\documentstyle[prd,eqsecnum,aps,graphicx,preprint]{revtex}

\begin{document}

\draft


\def \Tr {\mathop{\rm Tr}\limits}

\def \lg {\mathop{  \vbox{                  \hbox{${_{_<}}$}
                           \vspace{-.10 in }\hbox{${_{_>}}$}
                         }}\limits}

\def \gl {\mathop{  \vbox{                  \hbox{${_{_>}}$}
                           \vspace{-.10 in }\hbox{${_{_<}}$}
                         }}\limits}

\def \nin  {\mathop{ \hbox{ \hbox{\slash}
                            \hspace{-.18 in}  \hbox{$\in$} }
                       }\limits}

\def \bslp {\mathop{ \hbox{ \hbox{\slash}
                            \hspace{-.21 in}  \hbox{$p$} }
                       }\limits}

\def \bslk {\mathop{ \hbox{ \hbox{\slash}
                            \hspace{-.20 in}  \hbox{$k$} }
                       }\limits}

\def \bslm {\mathop{ \hbox{ \hbox{\slash}
                            \hspace{-.22 in}  \hbox{$m$} }
                       }\limits}

\def\REVTeX{REV\TeX}


\title{\Large{\bf{Infra-Red Asymptotic Dynamics of
                  Gauge Invariant Charged Fields:  
                  QED versus QCD}}}
 
\author{E. d'Emilio} 
\address{Dipartimento di Fisica, Universit\`a di Pisa {\rm{and}} I.N.F.N., Sezione di Pisa, I}
\address{ e-mail: {\tt demilio@difi.unipi.it}}

\vspace{-.8 cm}

\author{S. Miccich\`e}
\address{Department of Mathematical Sciences, Loughborough University, GB}
\address{ e-mail: {\tt S.Micciche@Lboro.ac.uk}}


\maketitle

\baselineskip = 13 pt


\vspace{-.6 cm}

\begin{abstract}

\vspace{-1 cm}

\noindent The freedom one has in constructing locally gauge invariant charged fields 
in gauge theories is analyzed in full detail and exploited to construct, in 
QED, an electron field whose two-point function $W(p)$, up to the fourth 
order in the coupling constant, is normalized with on-shell normalization 
conditions and is, nonetheless, infra-red finite; as a consequence the 
radiative corrections vanish on the mass shell $p^2=\mu^2$ and the free 
field singularity is dominant, although, in contrast to quantum field theories with mass
gap, the eigenvalue $\mu^2$ of the mass operator is not 
isolated. The same construction, carried out for the quark in QCD, is not 
sufficient for  cancellation of infra-red divergences to take place in the 
fourth order. The latter divergences, however, satisfy a simple 
factorization equation. We speculate on the scenario that could be drawn 
about infra-red asymptotic dynamics of QCD, should this factorization equation be
true in any order of perturbation theory.

\end{abstract}

\vspace{.2 in}


\narrowtext


\section{Introduction and Main Results}    \label{intro}

In ordinary Quantum Field Theory (QFT) with mass gap the notion of particle is recovered 
from that of interacting local field as a consequence of Infra-Red (IR) asymptotic dynamics: a 
near-mass-shell pole singularity in each of the momenta incoming any Green 
function (guaranteed in Lagrangian QFT's by the possibility of imposing 
on-shell normalization conditions on both mass and wave function 
renormalizations) ensures the existence of the Lehmann-Symanzik-Zimmerman asymptotic limit of the 
field $\cite{BLT}$. One is thus provided with a ordinary free Fock field, by means of
which an irreducible representation {\em \`a l\`a} Wigner of Poincar\'e 
group, sitting on an isolated mass hyperboloid, is in turn constructed. 
In this context the fact that the field/particle may or may not
carry quantum numbers associated with some unbroken global internal 
symmetry is irrelevant.

In gauge theories (we will always have in mind QED and QCD in continuum 
Minkowski 4-dimensional space-time with unbroken electric and colour 
charges) things go in a different way. Indeed, the issue is one about which, 
as yet, there is no general consensus. 

On the one side QED -- with the exception of its zero charge sector -- still 
is only a theory of inclusive  cross sections, in which all the theoretical 
set-up of quantum mechanics (states, observables, representation of 
symmetries and the like) has no satisfactory {\em explicit} representation, 
in spite of the general model-independent investigations $\cite{FMS,B1}$ 
that have  delimitated, so to speak, a possible battlefield: the battle is not yet won 
and one could, in a provocative way, summarize the situation by saying that
the question: ``what is an electron in QED'' is still open.

On the other side there is, in QCD, the problem of confinement of coloured 
gluons and quarks, about which there is even less to say. Many mechanisms
and criteria have been proposed over the years: some  
(as {\em e.g.} the Wilson loop area behaviour $\cite{W}$, or
the fundamental role of topology leading to the
dual Meissner effect $\cite{tH1}$) are so suggestive that have become common language;
others (the $1/(k^2)^2$ IR behavior of the full gluon propagator $\cite{BZ,B3PZ}$, $\cite{ADJS,AB}$, 
the quartet mechanism $\cite{KO}$ and the metric confinement $\cite{N1}$ both based on the existence of 
LSZ asymptotic limits for colour fields,
violation of asymptotic completeness $\cite{dEM1}$, 
the obstruction in the IR dressing due to Gribov ambiguities $\cite{LM}$,
and so many others that it would be impossible - and nonsensical - 
to quote them all here) do not share the same popularity, but time and again are
reconsidered and revived.
However, so far none of these criteria has led to a systematic and generally accepted
description of what confinement is.

Prudentially we regard confinement as a delicate, multi-faceted subject 
one can look at from different standpoints. We try here just to offer
a further standpoint, not necessarily in conflict with others, but 
endowed with the possibility of a sound mathematical verification based
on the only input of implementing in QCD the symmetries that we believe
relevant: local gauge invariance and Poincar\'e.

It is convenient to state the terms of the problem of the particle content 
of charged sectors in gauge theories within the framework of the Lagrangian 
approach. We shall also assume that all the fields entering the Lagrangian 
are local fields. These will be referred to as the basic fields of 
the model. Ref. $\cite{NO}$ gives in detail the local  covariant formulation
of the theory we shall rely on in the sequel. In particular the adjective 
``physical'' will be referred to the fields that commute with -- or to states 
that are annihilated by -- the Becchi--Rouet--Stora--Tyutin generator (the choice
of the local covariant formulation deserves a further comment: the fact that
manifest covariance is necessary to implement the renormalization procedure 
may be regarded upon as a technical complication; to our knowledge, however,
a proof of renormalizability is given only in this context $\cite{tH2}$: that 
is why we stick to it).

In this context it is convenient to distinguish four steps, all relevant in
designing the relationship between field and particle. We will try to
keep these steps as non-overlapping as possible:

(i) form of physical (composite) charged fields;

(ii) IR asymptotic dynamics;

(iii) existence of asymptotic limits and particle content;

(iv) $S$ matrix.

In this paper we will be mainly concerned with only (i) and (ii).

As for (i), it is well known that physical fields that are localized 
functions of the basic fields transform trivially ({\em i.e.} have zero 
charge) under any charge operator associated to a current obeying a  Gauss 
law: $j_\nu = \partial^\mu\,F_{\mu\nu}$. Indeed, in intuitive terms, thanks to the latter, the action of the 
charge on any field $\Phi$ takes the form
\begin{eqnarray}
                \delta \Phi = \lim_{R\to\infty} ~\left [ \int_{S_R}d{ S}_i\, { F}_{0 i}~,~ \Phi 
                                                 \right ]     
\label{GL}
\end{eqnarray}
where $S_R$ is the surface of the sphere of radius $R$ in 3-space. 
Therefore, if $\Phi$ is (or the fields in terms of which it is constructed 
are) smeared with functions of compact support, thanks to locality, 
$\delta\Phi$ vanishes for $R$ large enough. 
To avoid this the field  $\Phi$ must have a ``tail'' through the sphere at 
infinity in  Minkowski space (whether only in space-like or even in 
time-like directions is a subject to be taken up in the next section). 
In this sense,  as long as one is interested in  physical nontrivially 
charged fields, only  nonlocalized functions of the basic fields ought 
to be considered. 

Since the above statement has been given the status of a theorem $\cite{FPS}$, 
there is little to add and there is general agreement about it.

The theorem gives no hint, however, about the explicit form of such fields.
According to the terminology also recently used in Ref. $\cite{LM}$, such 
nonlocalized  functions will be  shortly referred to as ``dressed'' fields: 
a physical,  {\em interacting} electron should be dressed with a  cloud of 
photons, as well as with its own Coulomb field. 

Dirac $\cite{D}$ was the first to show, in an explicit way,  how the dressing could 
be done in order to endow an electron with its own Coulomb field. 
His aim was a quantization of QED that would involve only those degrees of 
freedom that actually contribute to the dynamic evolution of the system.
In retrospective, it does not sound as a surprise that he gave up the manifest 
covariance properties of the physical fields under Lorentz transformations: 
it was well known, after the Gupta-Bleuler formulation, that, even when 
restricting to the zero  charge  sector, such manifestly covariant 
formulations do involve indefinite  metric, {\em i.e} extra degrees  of 
freedom irrelevant to the dynamic evolution.

After Dirac other authors have investigated different ways of dressing the 
basic fields, with different motivations and with different aims.  
The list given by $\cite{BB}-\!\!\cite{HLM}$ only gives some references that 
are closer in spirit to the present article and, in any event, has no pretension
to completeness.
Ref. $\cite{LM}$ provides a much more comprehensive bibliography,
whereas $\cite{HLM}$ provides its updating.

On the same footing as Dirac, covariance is given up also in the model 
investigations of Steinmann $\cite{S1}$, who has the same aim as Dirac, and of Ref. $\cite{LM}$ 
and other works by the same group,
who instead think of the dressed fields as composite operators within the 
usual formulation of the gauge theory.   

The non-implementability of Lorentz boosts in the charged sectors  of QED is 
indeed, after the model independent investigations of $\cite{FMS,B1}$, taken for 
granted  to the point that, once the symmetry is broken by hand from the very beginning of 
the construction, no attempt is made to restore it. 
The only exception is provided, to our knowledge, by the attempts of one of
the present authors and collaborators $\cite{dEMfp,dEC}$. 
Needless to say, the effort of restoring Lorentz symmetry at some stage in 
our construction of dressed fields will be made also in the present paper. 
We have to state clearly that situating our results about QED within the 
general framework of $\cite{FMS,B1}$ is a non trivial subject, particularly because 
our results only concern some two-point functions and admittedly are, 
for now, incomplete.
It is true, on the other hand, that the results of the present and the 
following papers $\cite{dEMi}$ open the possibility of performing systematic 
model calculations that could help an explicit and exhaustive comparison: 
this is one of the several open, possibly not insurmountable, questions 
to be discussed in the conclusions.

Among the references we have cited, the work by Steinmann deserves a special
mention, for not only it has been close in spirit to ours along the years, but
it has been constantly inspiring. 
We feel it is not by chance that another part of Steinmann's and collaborator's work,
not immediately  connected with the problems specific to gauge 
theories, is invaluable to the approach presented here. Indeed, it turns out
that the usual Dyson expansion formula for the calculations of
Vacuum Expectation Values (VEV) of the type $\langle T(\cdots)\rangle$ is not sufficient for our purposes.
The composite fields we will introduce, will themselves be
$T^\pm$-ordered formal power series. So the calculation of their correlation functions
will demand the ability at computing -- in Perturbation Theory (PT) -- both Wightman functions and, 
more in general, multi-time-ordered
VEV's of the type $\langle T^\pm(\cdots)\cdots T^\pm(\cdots)\rangle$. Ref.s 
$\cite{O,S3,S4}$ exactly provide the algorithm for doing  all this.

Our attitude in the present paper is that we do not want to make any 
{\em a priori} assumption about IR asymptotic dynamics, with the exception 
of enforcing symmetries: local gauge, translations and Lorentz in particular. 
IR asymptotic dynamics should, hopefully, emerge by itself, {\em i.e.} only 
by our ability at calculating the near-mass-shell behaviour of correlation 
functions, once a particular gauge invariant charged field has been selected
within the framework of step (i) above.
In other words the main point is that (i) leaves a remarkable freedom and 
evidently any choice made  in selecting the form of physical charged fields 
may, and indeed does, affect the outcomes of (ii)-(iv).
Our work will, as a consequence, consist in exploiting all the freedom (i) 
leaves to see whether there exists a field with a near-mass-shell behaviour 
mild enough to enable one to eventually face point (iii) and (iv). 
In the case the motivations about the necessity of having fields
with a mild near-mass-shell behaviour should be recapitulated in more
intuitive and physical terms,  we have found the discussion given in 
$\cite{St}$ particularly sound.

Our expectation is that, playing, as we will do, twice the same game,
one should have different results in QED and QCD respectively. There is 
infact no point about the statement that the electron is not confined 
whereas, the quark should be so. 
Now, while in QED it is more or less generally accepted that the 
non-confinement of the electron should result in the existence of some kind 
of asymptotic (possibly not LSZ) limit, the way the confinement of quarks
and gluons should show up is less generally agreed upon: we have already
recalled some among the many, sometime conflicting, mechanisms that have been
proposed in the literature: needless to say, we will come out with
a mechanism different from all the others!

So, in order to directly compare QED and QCD,  we will construct  
``dressed electron'' 
$e(x)$ and  ``quark'' $q(x)$ fields (we could also construct the ``gluon'' 
$\cite{dEC}$, but the investigation of its behaviour in higher orders is better postponed to 
future work, for a 
comparison with its QED analogue would be less stringent: the photon has no 
charge) whose two-point functions up the fourth order in the coupling 
constant - the simplest place where a difference between QED and QCD may 
emerge -  have the following properties:

(1) they are independent of the gauge-fixing parameter;

(2) ultraviolet divergences brought about by the compositeness of the 
dressing are cured by a single renormalization constant introduced in the 
definition;
  
(3) on-shell normalization conditions can be imposed, in the IR regularized 
theory, on the single IR divergent graphs with two different outcomes.
   
(3a) In QED a complete cancellation of IR divergences takes place, and the
two-point function is given by
\begin{eqnarray}
              && W(p) =   \int d^4 x e^{i\,p\cdot(x-y)}\langle e(x) \overline{e}(y)\rangle   \quad ,
\label{WEdef}  \\
              && W(p) = W_0(p) + {\alpha\over \pi}\, W_1 + \left({\alpha\over \pi}\right)^2  W_2 + \cdots \quad ,
\label{WEres}
\end{eqnarray}
where
\begin{eqnarray}
                W_0(p) = (\bslp + \mu)\,(2\pi)\,\theta(p^0)\,\delta(p^2-\mu^2)                                 
\label{Wfree}
\end{eqnarray}
is the Wightman function of the free spinor field, whereas the higher order terms,
described by the two invariant functions $a_i(p^2/\mu^2)$ and $b_i(p^2/\mu^2)$:
\begin{eqnarray}
W_i(p)\, =\, \theta(p^0)\, \theta (p^2-\mu^2)\,{1\over \mu^2}\, (a_i \bslp + b_i \,\mu) 
\quad , \quad i \ge 1\quad ,
\label{W2par}
\end{eqnarray}
are given, to the first order, by

\begin{eqnarray}
         a_1 = {\mu^2\over 2\, p^2} \,\left (1 - {\mu^2\over p^2}\right )  \quad , \quad  b_1 = 0 \quad ,  
\label{W1} 
\end{eqnarray}
and, to the second order (whose full form is given in Section \ref{sec6}) have the
near-mass-shell, asymptotic form
\begin{eqnarray}
               a_2 \simeq 
                    {5\over 9}\, {\rm{r}} + {1\over 6} \, {\rm{r}}^2 \ln {\rm{r}} 
                    - {7\over 4}\, {\rm{r}}^2  + \cdots ~ , 
               \qquad
               b_2 \simeq 
                    -{7\over 36}\, {\rm{r}} - {1\over 6} \,{\rm{r}}^2 \ln {\rm{r}} 
                    + {5\over 24}\, {\rm{r}}^2  + \cdots ~, 
\label{W2} 
\end{eqnarray} 
with ${\rm{r}} = p^2/\mu^2 -1 \to 0 $.

(3b) In QCD, assuming dimensional regularization for IR divergences $-$ 
{\em i.e.}  $D=4\to 4 + 2\,\epsilon$ $\cite{GM,MS}$ $-$  the latter do not cancel, 
but obey the factorization equation   
\begin{eqnarray}
                 \epsilon \,{\partial \over \partial \epsilon} \, w_2(p,\epsilon) = 
                  + \left ({1\over 2\, \epsilon}\right )\, {11\over 6}\, C_{_A}  \, w_1(p,\epsilon) \quad ,  
\label{ede}
\end{eqnarray}
where $w = \sum (\alpha/\pi)^n\,w_n$ is defined by the amputation of 
the interacting part
\begin{eqnarray}
                && W_{ij}(p,\epsilon) = \int d^4 x e^{i\,p\cdot (x-y)}\langle q_i(x) \overline{q}_j(y)\rangle   \quad , 
\label{WQdef} \\
                && W_{ij}(p,\epsilon)  =   \delta_{ij}\,W_0(p)~+~     
                                         {i\over \bslp - \mu + i\,0}\,\,\delta_{ij}\, w(p,\epsilon)\,\, 
                                          {-i\over \bslp - \mu - i\,0} \quad . 
\label{wdef}
\end{eqnarray}
The first evident comment about the above results is that the game, 
played twice with the same rules,  gives two qualitatively different results.
It is sufficient, {\em per se},  to state that the IR asymptotic dynamics of
the two models is different  (this was expected), even in perturbation 
theory (possibly, this is a less  widespread belief).

Concerning $(\ref{WEres})$, although the result is that the singularity of 
the free field theory is not altered by the radiative corrections that 
vanish on the mass-shell, it is true that the mass hyperboloid $p^2=\mu^2$ 
is not isolated as in the mass gap case.
This result, expected on the basis of simple physical intuition, is in 
agreement with the observation made in Ref. $\cite{B2}$, where it has been pointed 
out that, in  the case of gauge theories, the particle content might be recovered at the cost of 
abandoning Wigner notion of an irreducible representation of Poincar\'e  group
sitting on an isolated mass hyperboloid.
The investigation of this point pertains the step (iii) above. 
We will not pursue it in this article.

Concerning the second result $(\ref{ede})$, we find it intriguing for two 
reasons. The first is that it is simple - we mean the factorization. 
The second is the occurrence of the celebrated ${11\over 6} \, C_{_A}$ factor, 
{\em with the plus sign}. 

We cannot therefore resist the temptation of  commenting on the consequences 
$(\ref{ede})$ would have, were it true in any order of PT.
In the latter case its integration would yield
\begin{eqnarray}
                && w(p,\epsilon) = e^{-{\alpha\over \pi}\,\Delta(\epsilon)}\,w(p) \,
                                \stackrel{\epsilon \to 0}{\longrightarrow} \, 0  \quad ,
\label{IRlim} \\
                && \Delta(\epsilon) = {11\over 6}\, C_{_A}\,{1\over 2\, \epsilon}  \quad ,
\label{Delta}
\end{eqnarray}
with $w(p)$ IR finite.

This hints at a different scenario, in which the Heisenberg ``quark'' field, 
{\em as a result of IR asymptotic dynamics}, is a free field not asymptotically, but at {\em any}
momentum $p$.

It may be useful to recall the example of the Faddeev-Popov (FP) ghost in QED: in that case 
the field is  free by construction and there is a factorization of 
correlation functions involving the ghost into a bunch of free  ghost two-point-functions times 
a connected correlation function only involving fields with zero ghost 
number. Could one say that the ghost number is confined?

In QCD, even if $(\ref{IRlim})$ were true, one could not immediately 
conclude, as in the case of local fields $\cite{SW}$ (we remind that dressed fields do
not share the locality property), that the $q_i(x)$ is a free field. 
Nonetheless a working hypothesis could be to check whether, as a consequence 
of IR asymptotic dynamics, the factorization of quark and gluon free 
two-point functions, out of connected correlation functions only involving 
colour singlets, does indeed take place.
The existence of asymptotic limit would, in the latter case, be by far 
simpler then in QED -- it would be trivial.

Of course, it is not necessary for the above scenario to take really place
that the function $\Delta(\epsilon)$ 
preserves, on possibly going from ($\ref{ede}$) to an exact result, the
specific form given by equation ($\ref{Delta}$) suggested by our fourth order
calculation. It might dress up even as a full
series in $\alpha$, provided that
$\Delta\to + \infty$ for $\epsilon \to 0^+$ carried
on holding. 

We are aware that, on extrapolating our result $(\ref{ede})$ to
$(\ref{IRlim})$, we have 
raised more questions  (all orders, gluon, IR asymptotics of 
many-point-functions) than we will answer in this article. 
But, in the framework we will set up, these questions do not seem to us 
prohibitively out  of the range of traditional and well established tools 
of QFT.

The paper is organized as follows.
In Section \ref{sec2} the freedom one has in dressing the basic fields is analyzed 
in detail on the level of classical fields. 
Section \ref{sec3} sets the stage for the calculation of quantum correlation 
functions: it is argued that an algorithm for computing VEV with  several time orderings,  {\em i.e.} of the type
$\langle T^{\pm}(\cdots)\cdots T^{\pm}(\cdots)\rangle$, 
is needed and the exhaustive work of Ostendorff and Steinmann $\cite{O,S3,S4}$, 
giving such an algorithm, is summarized. 
Section \ref{sec4} systematically explores in PT the lowest order 
of the two-point functions relative to the fields constructed in Section \ref{sec2},
and the full form of $W_1$, equation ($\ref{W1}$), is established.
Section \ref{sec6} gives a concise outlook of the fourth order calculations:
the full form of $W_2$, equation ($\ref{W2}$), is given together with a description of the way we follow
to calculate it and to obtain equation ($\ref{ede}$). The full derivation
of the latter results, as well as the proofs of their properties (1)-(3) 
above, are left for forthcoming papers.
In Section \ref{sec5} we give a retrospective of the construction we have done and pinpoint
the open problems that, in our opinion, most urgently should be faced in order to give 
the further, necessary support to such a construction.


\section{Classical Fields}    \label{sec2}

Let $\psi(x)$ denote a multiplet of Dirac fields transforming as the 
fundamental representation ${\cal R}$  of the colour group $SU(N)$ 
(the extension to whatever compact semi-simple Lie group being trivial). 
We shall denote by ${\bf A}_\mu(x) = t^a A_\mu^a(x)$ the Yang-Mills 
potentials. Here $t^a$, $a = 1 ,\cdots,\, N^2 - 1$, are the hermitian 
generators in ${\cal R}$, satisfying the commutation  relations 
$[t^a , t^b]= i\,f^{abc} \,t^c$, 
\mbox{$t_{il}^a \, t_{lj}^a = C_{_F} \delta_{ij}$}, 
\mbox{$C_{_F} = (N^2 - 1)/(2 N)$}; 
whereas the structure constants $f^{abc}$ are real, completely  antisymmetric
and obey \mbox{$f^{acd} f^{bcd} = C_{_A} \delta^{ab}$}, \mbox{$C_{_A} = N$}. 
The scalar and wedge products in ${\cal R}$ are accordingly defined by 
\mbox{${\bf A} \cdot {\bf B}=2 \Tr({\bf A}  {\bf B})$}, 
\mbox{${\bf A} \wedge {\bf B}= -i\,[{\bf A} \,,\, {\bf B}]$}.

It will be understood that the dynamics of the above fields is defined by the
Lagrangian ${\cal L}$ given, {\it ${\it e.g.}$}, in $\cite{NO}$, in which the
gauge-fixing term $- {\xi/ 2} \,(\partial {\bf{A}}) \cdot (\partial{\bf{A}})$ as 
well as the Faddeev--Popov ghosts have been introduced and the 
Becchi--Rouet--Stora--Tyutin symmetry is at work. 
All the fields in  ${\cal L}$ are assumed to be local fields.

Let ${\bf C}(x) = t^a\, C^a(x) \in {\cal R}$ be the FP ghost field, satisfying 
\begin{eqnarray}
{\bf C}(t,{\bf x}) \to 0~, \qquad 
{\rm{for}}~\vert{\bf x}\vert \to \infty~.                         
\label{Cdef}
\end{eqnarray}
We shall call local gauge transformations of $\psi$, $\overline{\psi}$ and 
${\bf{A}}_\mu$ the following:
\begin{eqnarray}
    \left.  \begin{array}{c}
                \delta {\bf A}_\mu = \partial_\mu {\bf C} + g\,{\bf A}_\mu \wedge {\bf C}~,  \\  
                \delta \psi = + ig\, {\bf C}\, \psi~,
                            \qquad 
                \delta \overline{\psi} = - ig\, \overline{\psi}\, {\bf C}~. 
    \end{array}  \right.    \label{BRSbasic}
\end{eqnarray}
Consider now the formal power series $\cite{S2}$: 
\begin{eqnarray}
      &&  V(y;f) = \sum_{{_{N=0}}}^{+\infty}\,(+ig)^{_N} \!\!
             \int \! d^4\eta_1 \!\cdots\! \int \! d^4\eta_{_N}  
            f_{_N}^{\nu_1\cdots\nu_{_N}}(y-\eta_1,\!\cdots\!,y-\eta_{_N}) 
            {\bf A}_{\nu_1}(\eta_1) \!\cdots\!{\bf A}_{\nu_N}(\eta_N)\,,         
\label{Vydef1} \\
      &&    V^\dagger(x;f) = \sum_{{_{M=0}}}^{+\infty}\,(-ig)^{_M}\!\!
             \int \! d^4\xi_1 \! \cdots \!\int \! d^4\xi_{_M}  
             f_{_M}^{\mu_1\cdots\mu_{_M}}(x-\xi_1,\!\cdots\!,x-\xi_{_M}) 
             {\bf A}_{\mu_{_M}}(\xi_{_M})\!\cdots\!  {\bf A}_{\mu_1}(\xi_1) 
\label{Vxdef1}                         
\end{eqnarray}
where the terms $M , N = 0$ are by definition $1$.

We claim that one can choose {\em real} kernel functions $f$'s such that $V$ 
and $V^\dagger$ transform  under $(\ref{BRSbasic})$ according to
\begin{eqnarray}
         \delta V   =+ i\,g~{\bf C}~V~, \qquad 
         \delta V^\dagger =- i\,g~ V^\dagger~ {\bf C}~. \label{BRSV}
\end{eqnarray}
Before we proceed to enforce the transformation properties $(\ref{BRSV})$, 
two comments are in order about the multiple convolutions displayed in 
$(\ref{Vydef1})$ and $(\ref{Vxdef1})$.

(i) The first is that they are mandatory if one is interested, as we are, in 
obtaining translation covariant solutions to $(\ref{BRSV})$.

(ii) The second is that the convolutions extending to the whole Minkowski 
space explicitly expose the fact that $V$ and $V^\dagger$ may be 
non-localized functions of the basic local fields ${\bf{A}}$, provided the 
support of the $f$'s is suitably chosen. In view of the discussion about 
$(\ref{GL})$, this is quite welcome because we are aiming at constructing 
locally gauge invariant fields that carry nontrivial global colour numbers: 
indeed, concerning local gauge transformations, once $(\ref{BRSV})$ are 
satisfied, the spinor fields:
\begin{eqnarray}
       \Psi_f(x)=V^{\dagger}(x;f)\psi(x)  \quad ,\quad
      \overline{\Psi}_f(y)=\overline{\psi}(y)V(y;f)  
\label{PSIdef1}
\end{eqnarray}
are obviously invariant under $(\ref{BRSbasic})$ while they transform as 
${\cal R}$ and ${\cal \overline{R}}$ when ${\bf{C}}$ is not chosen according
to $(\ref{Cdef})$, but is constant with respect to $x$.

Let us go back to enforcing $(\ref{BRSV})$. Steinmann has faced this problem 
in Ref. $\cite{S2}$.
He assumes that, on introducing $(\ref{BRSbasic})$ into $(\ref{Vydef1})$ and 
$(\ref{Vxdef1})$, the derivatives can be reversed by parts. 
While this can be justified for space derivatives, thanks to the boundary
conditions $(\ref{Cdef})$ on the ghost, the thing is less justifiable for the
time derivatives, as one has no {\em a priori} control on asymptotic behaviour
in time. In electrodynamics there is a way out: since the ghost is free, one 
can choose suitable solutions of the d'Alambert equation $\cite{dEMfp}$ that 
justify the neglect of boundary terms. In the non-abelian case the problem is
there: we shall, as in $\cite{S2}$, just ignore it, recalling however the 
statement (1) of the introduction that, in the case  of quantum fields, we 
will be able to prove the $\xi$-independence of correlation functions.

With this {\em proviso}, Steinmann has shown that the requirement that 
$(\ref{BRSV})$ be satisfied by $(\ref{Vydef1})$ and $(\ref{Vxdef1})$ order by 
order in $g$ leads to a linear inhomogeneous recursive system for the 
$f$'s. The Fourier transforms of the first of the equations he gives is:
\begin{eqnarray}
       k_\nu~\hat{f}_1^{\nu} (k)~ = ~i~,                                                    
\label{STFourier1}
\end{eqnarray}
whereas the $f$ with $N>1$ arguments is determined in terms of the $f$ with
$N-1$ arguments by 
\begin{eqnarray}
         &&  \left \{
             \matrix{
              (k_1)_{\nu_{1}} {\hat f}_{{_N}}^{\nu_{1} \cdots \nu_{{_N}}}
                    (k_{1},\cdots, k_{{_N}}) = i\,                         
                    [{\hat f}_{{_{N-1}}}^{\nu_{2} \cdots \nu_{{_N}}} 
                    (k_{2},\cdots, k_{{_N}})  
                     - {\hat f}_{{_{N-1}}}^{\nu_{2} \cdots \nu_{{_N}}} 
                     (k_{1}+k_{2},k_{3},\cdots, k_{{_N}})]~,  \hfill    \cr
                       \cr
              (k_{\alpha})_{\nu_{\alpha}} {\hat f}_{{_N}}^{\nu_{1} \cdots 
                    \nu_{\alpha} \cdots \nu_{{_N}}} 
                    (k_{1},\cdots,k_{\alpha}, \cdots , k_{_N}) = \hfill \cr
                                                                        \cr
     \qquad \qquad   
              +i \,\bigl[{\hat f}_{{_{N-1}}}^{\nu_{1} \cdots \nu_{\alpha -1} 
                   \nu_{\alpha +1} \cdots \nu_{_N}} 
                   (k_{1},\cdots, k_{\alpha -2}, k_{\alpha -1}+
                   k_{\alpha}, k_{\alpha +1},\cdots,k_{_N})~+ \hfill    \cr
                       \cr
     \qquad \qquad   
              ~~~-~{\hat f}_{{_{N-1}}}^{\nu_{1}\cdots \nu_{\alpha- 1} 
                   \nu_{\alpha +1}\cdots\nu_{_N}} 
                   (k_{1}, \cdots, k_{\alpha -1}, k_{\alpha}+ k_{\alpha +1}, 
                   k_{\alpha +2},\cdots,k_{_N})\bigr]~,       \hfill    \cr
                        \cr
                   (k_{{_N}})_{\nu_{{_N}}} {\hat f}_{{_N}}^{\nu_{1} \cdots 
                   \nu_{{_N}}}
                   (k_{1}, \cdots, k_{{_N}}) = 
                   i\,{\hat{f}}_{{_{N- 1}}}^{\nu_{1} \cdots \nu_{{_{N-1}}}}
               (k_{1},\cdots, k_{{_{N-2}}}, k_{{_{N-1}}}+k_{{_N}})~.\hfill\cr
             } \right.                                                               
\label{STFourierJ} 
\end{eqnarray}
with ${2 \le \alpha \le N-1}$. We also take from $\cite{S2}$ that the solutions of $(\ref{STFourier1})$ and $(\ref{STFourierJ})$, that for any integer ${N}$ satisfy:
\begin{eqnarray}
         &&\sum_{_{J = 0}}^{{_N}}(-1)^{_J}~
           \hat{f}_{_J}^{~\nu_1 \cdots \nu_{_J}} (k_1,\cdots, k_{_J})~ \hat{f}_{{_{N - J}}}^{~\nu_{{_N}} \cdots 
           \nu_{{_{J+1}}}} (k_{_N},\cdots, k_{_{J + 1}})~=~0 \quad ,
\label{UNIa}   \\
         &&\sum_{_{J = 0}}^{{_N}}(-1)^{{_{N - J}}}~
           \hat{f}_{_J}^{~\nu_{_J} \cdots \nu_{1}} (k_{_J},\cdots, k_1)~ \hat{f}_{{_{N - J}}}^{~\nu_{_{J +1}} \cdots 
           \nu_{{_N}}} (k_{{_{J + 1}}}, \cdots, k_{_N})~=~0  \quad ,
\label{UNIb}
\end{eqnarray} 
give rise to unitary series \mbox{$V(x;f)~V^\dagger(x;f) = V^\dagger(x;f)~V(x;f) = 1$}.

Let us first focus on $(\ref{STFourier1})$. A family of solutions to this 
equation that also satisfies $(\ref{UNIa})$ and $(\ref{UNIb})$ isa
\begin{eqnarray}
      \hat f_1^\nu (k;c) = 
      i\,n^\nu~{1\over 2}\left({1+c\over n\cdot k - i\,0}+
      {1-c\over n\cdot k + i\,0}\right ) \quad ,
\label{f1c}
\end{eqnarray}
where $c$ is a real parameter and $n^\nu$ is a 4-vector that we leave, for 
the moment, unspecified.
Two particular solutions from $(\ref{f1c})$ are
\begin{eqnarray}
      && \hat f_{+\,1}^\nu(k;n) = {i\,n^\nu\over n\cdot k - i\,0}  \label{f1+} \quad , \\
      && \hat f_{-\,1}^\nu(k;n) = {i\,n^\nu\over n\cdot k + i\,0}  \label{f1-} \quad .
\end{eqnarray}
It can be verified that the two following sets of functions $\hat{f}_{+{_N}}$ 
and  $\hat{f}_{-{_N}}$, given by
\begin{eqnarray}
               \hat{f}_{\pm{_N}}^{\nu_1\cdots\nu_N}(k_1,\cdots,k_N;n) =  
                {{i\, n^{\nu_1}} \over {n \cdot (k_1 + \cdots + k_{{_N}}) \mp i\, 0}}~
                \cdots {{i\, n^{\nu_{_N}}} \over {n \cdot k_{{_N}} \mp i\, 0}}  \label{-ie}   
\end{eqnarray}
separately satisfy all equations $(\ref{STFourierJ})$-$(\ref{UNIb})$. 
These solutions also fulfil the factorization property
\begin{eqnarray}
  && \sum_{\rm perm} \, 
     \hat{f}_{\pm{_N}}^{\nu_1\cdots\nu_N}(k_1,\cdots,k_N;n) ={1\over N!}\,
    \hat{f}_{\pm{_1}}^{~\nu_1}(k_1;n)\cdots\hat {f}_{\pm{_1}}^{~\nu_{_N}}
    (k_{_N};n)        \label{eik}
\end{eqnarray}
well known as eikonal identity, as well as the $n-$reflection exchange 
relation
\begin{eqnarray}
      \hat{f}_{\pm{_N}}^{\nu_1\cdots\nu_N}(k_1,\cdots,k_N;n) =      
      \hat{f}_{\mp{_N}}^{\nu_1\cdots\nu_N}(k_1,\cdots,k_N;-n)  \quad .     
      \label{f-m}
\end{eqnarray} 

We will also need a third set of solutions, that extend to higher orders the 
lowest order solution obtained by setting $c=0$ in $(\ref{f1c})$:
\begin{eqnarray}
     \hat f_{0\,1}^\nu (k;n) = 
     {i\,n^\nu~\over 2}\left({1\over n\cdot k - i\,0}+
     {1\over n\cdot k + i\,0}\right )
\label{f10}
\end{eqnarray}
with Principal Value prescription (PV) for the $n\cdot k$ denominator. 
This is evidently connected with the problem of exposing a family of 
solutions that interpolates between $\hat{f}_{+{_N}}$ and  $\hat{f}_{-{_N}}$.
We have found that, with $n^\nu$ kept fixed and even after imposing the 
unitarity constraints $(\ref{UNIa})$ and $(\ref{UNIb})$, the higher the ${N}$
the higher the number of complex parameters due to the occurrence of 
Poincar\'e-Bertrand terms.
However, if also the eikonal identity $(\ref{eik})$ is enforced, the 
interpolating family only depends on the real parameter $c$ appearing in 
$(\ref{f1c})$. 
Just to give a flavour of the thing, it is found that
\begin{eqnarray}
               && \hat f^{\nu_1\nu_2}_2 (k_1,k_2;c) =  
                  \hat f_1^{\nu_1} (k_1+k_2;c) ~ \hat f_1^{\nu_2} (k_2;c) ~  
                     +{\pi^2\over 2}(1-c^2)\,
                   n^{\nu_1}\,n^{\nu_2} \delta(n\cdot k_1)\,\delta(n\cdot k_2) \quad .
\end{eqnarray}
We have explicitly found up to $f_4(k_1,\cdots,k_4;c)$ and we also have a 
guess about $f_N(k_1,\cdots,k_N;c)$ for generic ${N}$. 
But, for the sake of conciseness we will no longer elaborate on this topic, 
also because higher orders will not be needed in the perturbative 
calculations we will perform in later sections.
The important for the sequel is that there exists a solution, denoted by
$\hat f^{{\nu_1}\cdots{\nu_N}}_{0{_N}}(k_1,\cdots,k_N;n)$, 
that extends $(\ref{f10})$ to any order ${N}$. In connection with 
$(\ref{f-m})$, note that the solution $\hat f_0$, in addition to satisfying 
the eikonal identity, is also invariant under $n-$reflection
\begin{eqnarray}
                \hat{f}_{0{_N}}^{\nu_1\cdots\nu_N}(k_1,\cdots,k_N;n) =      
                \hat{f}_{0{_N}}^{\nu_1\cdots\nu_N}(k_1,\cdots,k_N;-n)    \quad .   
\label{f0-m}
\end{eqnarray}

The relationship between the present approach and other ones 
$\cite{LM,M,S1,S2}$, can now be clarified.

Consider, to this purpose,  $V_-(y;n)$, {\em i.e} the $V$ obtained by 
inserting the solution $\hat f_{-{_N}}$ ({\em i.e} $(\ref{f1-})$ and 
$(\ref{-ie})$ with the $-$ sign) into $(\ref{Vydef1})$. It is useful to 
represent all the denominators in $\hat f_{-{_N}}$ by means of the 
one-parameter integral representation 
\mbox{  $(b + i \,0)^{-1} = -i\int_{0}^{+\infty}d \omega\,
{\rm{exp}}\,[i \omega\,(b + i \, 0)]$}. 
In this way it is possible to explicitly perform the $d_4 k_j$ 
integrations in the anti-Fourier transform of the $\hat f$'s. 
These integrations give rise to 
$\delta_4(x - \eta_j + \sum_i \omega_i)$ 
that allow, in turn, for the elimination of the $d^4 \eta_j$ integrations in 
$(\ref{Vydef1})$. Some further obvious manipulations convert $(\ref{Vydef1})$ 
into
\begin{eqnarray}
      & V_-(y;n) &= \sum_{{_{N = 0}}}^\infty \, ( i \,g)^{_N} \!
         \int_0^{+\infty} \!\! d \omega_1 \,\cdots 
         \int_{\omega_{_{N - 1}}}^{+\infty} \!\! d \omega_{_N}
           \, n \cdot {\bf{A}} (y - n\, \omega_1)\cdots n \cdot 
         {\bf{A}} (y - n\, \omega_{_N}) ~=       \nonumber \\
      & & = {\cal P^{+}}~ \exp \left [
        {{ i\,g \int_0^{+\infty} \!d \omega\,n 
         \cdot {\bf{A}}(y - n\, \omega)}}\right] \quad .              
\label{V-string}
\end{eqnarray}
The r.h.s. of the above formula is the usual definition of the path-ordering 
symbol ${\cal P^{+}}$. If $n$ is chosen to be a space-like vector, the above 
representation clarifies that $V_-$  is nothing but a rectilinear string 
operator ${\em \acute{a}\,l\acute{a}}$ Mandelstam $\cite{M}$ extending to 
space-like infinity. The case of $n$ space-like may serve also to accommodate Buchholz case 
$\cite{B3}$. For this reason we will generically refer to all the  $V$ and $V^\dagger$ 
operators as to the string operators, regardless of whether $n$ is space-like
or time-like.

In the same way one finds that
\begin{eqnarray}
   V^\dagger_+(x;n) = 
   {\cal P^{+}} \exp\left [{\displaystyle{- i\,g \int_0^{+\infty}\! d \omega\,n 
   \cdot {\bf{A}}(x + n \omega)}}\right]              
\label{V+string}
\end{eqnarray}
(again a ${\cal P^{+}}$ for the order of the $n\cdot{\bf A}$ factors in 
$V^\dagger$ is reversed with respect to $V$).

It is now convenient to introduce the decomposition of the Minkowski 4-space 
${\cal M}_4$ into the future and past light cones, and their complement:
\begin{equation}
                {\cal M}_4 = {\cal C}_+\cup {\cal C}_0 \cup {\cal C}_-\label{Mink}
\end{equation}
and in the sequel, referring to the above decomposition, the indices $\sigma$ and $\tau$ will always take the values $\pm,\,0$.

Suppose now that, we choose $n\in {\cal C}_{\pm}$, {\em i.e.} in the 
future/past light-cone. 
Then the statement that respectively  $n_0\gl0$ is Lorentz invariant, 
whence also 
$x^0 - \omega_i\,n^0 = \tau_i \lg \tau_{i-1} \lg \cdots \lg \tau^0 = x^0$.
In view of this, $(\ref{V-string})$  can be written using the $T^\pm$ chronological ordering symbols
\begin{eqnarray}
               && V_-(x;n) =
                   { T^{\pm}} \exp\left [{\displaystyle
               {+i\,g \int_0^{+\infty}\! d \omega\,n \cdot {\bf{A}}(x - n\, \omega)}}\right ] \quad ,             
                ~~n\in{\cal C}_{\pm}  \quad , 
\label{V-stringt}
\end{eqnarray}
and likewise for $(\ref{V+string})$
\begin{eqnarray}
               && V^\dagger_+(x;n) = 
                    T^{\pm} \exp\left [{\displaystyle
                    {-i\,g \int_0^{+\infty}\! d \omega\,n \cdot {\bf{A}}(x + n\, \omega)}}\right ] \quad,             
                    ~~n\in{\cal C}_{\pm}  \quad . 
\label{V+stringt}
\end{eqnarray}
So far this is no big difference: the ordering operators, either 
${\cal P^{\pm}}$ or ${ T^{\pm}}$, only order the colour matrices 
$t^{a_1},\cdots,t^{a_N}$ in the $N$-th term of the above series, whereas the
fields $A^{a_1}_{\nu_1},\cdots,A^{a_N}_{\nu_N}$, inasmuch as classical fields, 
are not sensitive to this ordering. 
In the case of classical electrodynamics -- $t^a\to 1$ -- such operators are 
simply useless.
The role of the $T^{\pm}$ ordering will instead become crucial when we will 
keep it in the definition of the quantum Heisenberg operators.

We have also to consider string operators in which the string vector $n$ is 
chosen space-like. In this case the difference between the arguments of two 
neighbouring $A$'s is space-like, so only the colour matrices are 
sensitive to the ordering, whereas even the Heisenberg fields of the quantum 
case commute with one another, due to locality.
The solution we will consider for $n \in {\cal C}_0$ are 
$V_0(x;n)$ and $V^\dagger_0(x;n)$,  {\em i.e.} the ones corresponding to the 
solution $\hat f_{0_{N}}$ that extends 
$(\ref{f10})$ and fulfils the $n$-reflection invariance property $(\ref{f0-m})$.

Let us now introduce the characteristic functions
\begin{eqnarray}
  \chi_\sigma(n) =  \left \{ \begin{array}{lll}
                                          1 & {\rm if} & n\in {\cal C}_{\sigma} \\  
                                          0 & ~        & {\rm{otherwise}} 
                 \end{array} \right. \quad,
    \qquad \qquad 
  \sigma = \pm 1, \, 0                                
\label{chi}                                                                        
\end{eqnarray}
and correspondingly the fields
\begin{eqnarray}
    &&  \left. \begin{array}{l}
             \Psi_\pm(x;n) = \chi_\pm(n) ~~T^\pm\bigl[ V^\dagger_\pm (x;n) \, \psi(x)\bigr] \quad ,  \\
             \Psi_0  (x;n) = \chi_0  (n) ~~ V^\dagger_0   (x;n) \, \psi(x) \quad ,
        \end{array} \right.      \label{psi+-0} \\
    &&  \left. \begin{array}{l}
             \overline \Psi_\pm(x;n) = \chi_\pm(n)~~  T^\pm \bigl [\,\overline \psi(x) \, V_\pm(x;n) \bigr] \quad , \\
             \overline \Psi_0  (x;n) = \chi_0  (n) ~~ \overline \psi(x) \, V_0  (x;n) \quad . 
        \end{array} \right.       \label{psibar+-0}
\end{eqnarray}
These fields fulfil the Dirac conjugation properties
\begin{eqnarray}
       \left. \begin{array}{l}
             \overline {\Psi_\pm}(x;n) = \overline \Psi_\mp(x;-n) \quad ,    \\
             \overline {\Psi_0}  (x;n) = ~\overline \Psi_0  (x,-n) \quad ,
       \end{array} \right.       \label{DiracConj+-0} 
\end{eqnarray}
that follow from the $n$-reflection properties $(\ref{f-m})$ and $(\ref{f0-m})$.
As a consequence the composite fields
\begin{eqnarray}
       \left. \begin{array}{l}
             \Psi(x;n) =  z_+\,\Psi_+ + z_-\,\Psi_- +  z_0\, \Psi_0 ,     \\
             \overline \Psi(x;n) = z_-\, \overline \Psi_+ + z_+\,\overline \Psi_- + z_0\, \overline \Psi_0 
       \end{array} \right.      \label{zitazeta}  
\end{eqnarray}
(with the complex constants $z$'s satisfying 
$\overline z_\pm = z _\mp$, $\overline z_0 = z_0$, and to be specified later, for the quantum fields, when effecting renormalization) 
satisfy the Dirac conjugation relation
\begin{eqnarray}
\overline \Psi(x; n) = \Psi(x;-n)^\dagger \, \gamma_0          
\label{DiracConjx}
\end{eqnarray}
that, in Fourier transform, takes the form
\begin{eqnarray}
\overline {\hat \Psi}(p; n) = {\hat {\overline\Psi}}(-p;-n) \quad .           
\label{DiracConjp}
\end{eqnarray}

All the constructions done so far, to go from $\psi$ to $\Psi$, can be crudely 
summarized in this way: 
one has traded the gauge-variance of $\psi$ for the dependence of $\Psi$ on 
the string vector $n$.
We will refer to this fact as a breaking, put in by hand, of the original 
Lorentz symmetry -- an unpleasant feature one would like to get rid of. 
We dedicate the rest of this section to give a heuristic description of how we 
will try to accomplish this task.

The Dirac equation for the ordinary $\psi$ in linear covariant gauges is first
converted into the equation of motion for ${\Psi}(x;n)$. 
We write it in momentum representation:
\begin{eqnarray}
      && (\bslp - \mu) \, \hat{\Psi}(p;n) =  
          g~\gamma_\alpha\,t^a \int d_4 k \sum_\sigma\,z_\sigma\,
                    T_\sigma^{ \alpha\beta}(k;n)~ 
                   {{A}}_\beta^a(k) \, \hat{\Psi}_\sigma(p- k;n)   =                      \nonumber \\
      && \hspace{1.15 in} =  g~\gamma_\alpha\,t^a\, Q^{a\alpha}(p;n)   \quad ,                        
\label{mDirac} 
\end{eqnarray}
where the index $\sigma$ refers to the decomposition of $\Psi$ with respect to
the light-cone of $n$, equation
$(\ref{psi+-0})$. Accordingly, the projectors $T$ are given by
\begin{eqnarray}
          \left. \begin{array}{l}
                 T^{ \alpha\beta}_\mp(k;n) = 
                 g^{\alpha \beta} - \displaystyle{  {{k^\alpha n^\beta}~ \over {n\cdot k \pm i\, 0}} }  ,   \\
                     \\
                 T^{ \alpha\beta}_0(k;n) = 
                 g^{ \alpha\beta} - \displaystyle{  {\rm PV}~{ {{k^\alpha n^\beta}}\over {n\cdot k }}  }
         \end{array} \right.      \label{Ts} 
\end{eqnarray}
and satisfy
\begin{eqnarray}
         n_\alpha~T^{ \alpha\beta}_\sigma(k;n) = 0~, 
            \qquad 
         T^{ \alpha\beta}_\sigma(k;n)~k_\beta = 0 ~.   \label{mTk}
\end{eqnarray}
Thanks to the second of equations $(\ref{mTk})$, the longitudinal degrees of freedom of $A_\beta^a$ are expected to
decouple. Thanks to the first of equations $(\ref{mTk})$, the vector field to which $\Psi$ 
is coupled is ${\cal A}^{a\alpha}=T^{\alpha\beta}\,A^a_\beta$ that satisfies 
$n\cdot {\cal A}^a=0$. 

Were it not for the subtleties due to the $\pm i\,0$ prescriptions
({\em i.e.} to the light cone decomposition of the field with respect to $n$), 
this formally is the equation satisfied by the Dirac field in the axial gauge. 
One could try to take this as a substitute of the ordinary Dirac equation in 
linear covariant gauges and $\Psi(x;n)$ (with $n$, as in a gauge-fixing, 
chosen once for all) as the variable substituting $\psi$ and in terms of which 
to attempt a gauge-invariant formulation of the theory -- much in the spirit of 
$\cite{D,M,S1}$.

We will not take this attitude. We will continue to think of $\Psi(x;n)$ as a 
composite field in a theory where $\psi$ and $A_\alpha^a$ play the role of 
basic dynamic variables. 

This point of view leaves open the possibility of choosing different $n$'s for 
different $\Psi$'s.
More clearly, we want to leave open the possibility of computing quantum 
correlation functions of the type $\langle\Psi(x;m)\overline\Psi(y;n)\rangle$,
in which any field has its own string and with no restriction
on whether both $m$ and $n$ are taken either time-like or space-like.

This also is the point where we can explain how we will recover the lost 
Lorentz symmetry.
We will discuss about the possibility of taking the limit
\begin{eqnarray}
                n\to p              
\label{ntop}
\end{eqnarray} 
in equation $(\ref{mDirac})$. 

A serious warning about this limit is that 
its very existence is far from being trivial: we will give some positive 
evidence in favour of it only in the case of quantum fields in Section \ref{sec4}. 

For now we will just forget about any mathematical rigor and assume its
existence: this enables us to draw some conclusions and formulate some 
expectations about quantum fields.

The first consideration about $(\ref{ntop})$ is that it does not mess up the 
Dirac conjugation properties of $\Psi$, as evident from equation 
$(\ref{DiracConjp})$. 

Let us then call
\begin{eqnarray}
\hat q(p) = \hat \Psi(p;p) \quad .    
\label{defq}
\end{eqnarray} 
Then, by setting $n = p$ in $(\ref{mDirac})$, one obtains:
\begin{eqnarray}
     && (\bslp - \mu) \,\hat{q}(p) = 
        g\,\gamma_\alpha\,t^a \int d_4 k \sum_\sigma\,z_\sigma\,
                 T_\sigma^{\alpha\beta}(k;p)\, 
              {\hat{A}}_\beta^a(k) \, \hat{\Psi}_\sigma(p - k;p)                     
\label{eq Dirac2} 
\end{eqnarray}
that makes evident why we have kept our point of view: 
differently from $\hat \Psi(p;n)$, the field $\hat q$  may exist only as a 
composite field: in the r.h.s. of the above equation the  $\hat \Psi$ appears 
with two different values of its arguments, so the field $\hat q$ does not 
satisfy a closed equation.
For $\hat q$, as already for $\hat \Psi(p;n)$, it is expected that the 
unphysical degrees of freedom of $A_\alpha^a$ decouple: the second of equations 
$(\ref{mTk})$ still applies. 

But this is not the end of the story.
If, according to a well known argument, the near-mass-shell behaviour of the 
field is driven by the classical currents  responsible for the interaction 
with  soft gluons/photons, we can make a guess about it by operating the 
replacement $\gamma^\alpha \to \mu\, p^\alpha/p\cdot k$
within the integration in the r.h.s. of $(\ref{eq Dirac2})$. 
It is then seen that, thanks now to the first of equations $(\ref{mTk})$ with $n=p$, 
also the classical currents decouple and no longer drive the asymptotic IR dynamics
of $q$. As a result, the near-mass-shell behaviour of the field $q$ we have 
defined should be at least milder than that of both the gauge-variant $\psi$ 
and the $n$-dependent $\Psi$.

The observation above, finally, clarifies why we have constructed strings 
allowing for the choice of a time-like vector: in the classical currents 
the momentum is close to the mass-shell: $p^2\simeq\mu^2>0$.

All these expectations for the quantum fields will find confirmation in the 
following sections.
This means that we will give meaning, to some extent, to the heuristic formula
\begin{eqnarray}
q(x) = \int {d^4p\over (2\pi)^4}\, e^{-i\,p\cdot x}\, 
\left [ \int d^4y\,e^{i\,p\cdot y}\,\Psi(y;n) \right ]_{n=p}
\end{eqnarray}
with $\Psi(y;n)$ given by $(\ref{zitazeta})$.
The utility of this formula is to clarify that the kind of delocalization 
involved in $q(x)$ is by far more complicated than that, recalled in connection with 
($\ref{GL}$), of a field with a 
``tail'' going to  infinity along a string that
is rectilinear in coordinate representation, as is the case for $\Psi(y;n)$.
In pictorial terms the strings contributing to $q(x)$ are spread out all
over $x$-space: this happens when a string, rectilinear in $p$-space,
is integrated upon with $\exp(-i\,p\cdot x)$ as weighting factor.
The field $q(x)$ thus rather resembles a kind of space-time candy-sugar 
cloud centered at $x$.


\section{Perturbation theory for Quantum Fields}    \label{sec3}

The present section is devoted to set up diagrammatic rules for the 
calculation, in perturbation theory, of the correlation functions of the 
quantum gauge invariant charged fields we have sketched in Section \ref{sec2}.

We define the quantum field corresponding to $(\ref{zitazeta})$ in the 
following way:
\begin{eqnarray}
             && \Psi(x;m) = \int \!{d^4p\over (2\pi)^4}~ e^{-i\,p\cdot x}\, 
                                  \Biggl \{\sum_{M=0}^\infty (-ig)^M \,~t^{a_M}\cdots t^{a_1}\, 
                                  \prod_{j=1}^M  \int {d^4k_j\over (2\pi)^4}
                                  ~\times           \label{PsiPT} \\
             &&         \times \, \sum_{\sigma=\pm,0}\, \chi_\sigma(m)\, \zeta_\sigma^{1/2} \, 
                                  \langle V^\dagger_\sigma \rangle ^{-1} \,
                                  \hat f_{\sigma{_M}}^{\mu_1\cdots\mu_M}(k_1,\cdots,k_M;m) \,\, T^\sigma \,
                                  \Bigl [\hat A_{\mu_M}^{a_M}(k_M)\cdots \hat A_{\mu_1}^{a_1}(k_1) \,
                                  \hat{\psi} \Bigl(p-\sum_{j=1}^M k_j\Bigr)\Bigr] \Biggr\}.  \nonumber  
\end{eqnarray}
In the above formula the time-ordering operators $T^\pm$ and the identity 
operator $T^0 = 1$ act on the Heisenberg fields in the square bracket. Moreover
$ \zeta_+ = \zeta_-$ and $\zeta_0$
will play the role of real renormalization constants, introduced to take care of
the compositeness of $\Psi$.
In addition, also the factors $\langle V^\dagger_\sigma \rangle^{-1}$
are constants whose values will be fixed later, when their necessity to avert 
some ill-defined one particle reducible graphs  will be realized. 
For now, what is needed to know is that the $\zeta_\sigma$ and the
$\langle V^\dagger_\sigma \rangle^{-1}$ have the right conjugation properties
such that 
$\zeta_\sigma\,\langle V^\dagger_\sigma \rangle^{-1}$ 
can be identified with 
the $z_\sigma$ of equation ($\ref{zitazeta})$: in this way $\overline \Psi(x;m)$ is, in 
turn,  obtained by taking the  straightforward  Dirac conjugate of 
($\ref{PsiPT}$).

It should be finally noted that the structure of formula $(\ref{PsiPT})$ is 
slightly different from $(\ref{zitazeta})$.
In fact, in the latter case one can recognize the time-ordering of the fields 
only after performing, as we have done in Section \ref{sec2}, the $d^4\xi_j$ 
integrations of $(\ref{Vxdef1})$. Here, instead, the $d^4k_j$
integration involving the $f$'s, that are in turn responsible for
this ordering, are indicated but not yet performed: 
the $T^{\pm,0}$ are there simply by definition.

The light-cone decomposition of the field with respect to $m$ -- 
the second line of the above 
formula -- makes it evident that, depending on the choice of the string vector 
$m$ relative to any single field, one  must be able to compute VEV's of the type 
$\langle T^{\sigma_n}(\cdots)\cdots T^{\sigma_1}(\cdots)\rangle$,
with $\sigma_i=\pm,\,0$. This observation entails that the usual Dyson perturbation theory formula for the development of
one single $T$-ordered product is not sufficient to our purposes. 

An extension of Dyson algorithm is therefore needed and,
fortunately for us, such an extension is already available, thanks to the
work of Ostendorff $\cite{O}$ and Steinmann $\cite{S3}$. 

We recapitulate their results for the reader's convenience (reporting more or 
less {\em verbatim} the content of the appendix of Ref. $\cite{S4}$).

Let us denote by $X=\{x_1,\cdots,x_r\}$ a set of 4-vectors and $\Phi$ stand for any basic field ($A_\alpha^a,\,\psi$ {\em etc}.)
of interest for us. Let also $T^\sigma(X)$ denote the corresponding product of 
the fields $\Phi(x_1)\cdots \Phi(x_r)$. In the multi-time-ordered vacuum expectation value
\begin{eqnarray}
                W(X_n,\sigma_n |~...~| X_2,\sigma_2  | X_1,\sigma_1)=\langle T^{\sigma_n}(X_n)\cdots T^{\sigma_1}(X_1)\rangle
\end{eqnarray}
any $\sigma_i$ may take the value $\pm$ only
(the case $T^0=1$ of no ordering will be included
later). The perturbative contribution to order $g^N$ to $W$ is obtained as follows. 

$\bullet$ {\em{Graphs}}: All the graphs with $\sum\, r_i$ external points and a number
of internal points suitable to match the order $N$ in PT are drawn.

$\bullet$ {\em{Partitions}}: Each of the above graphs is partitioned in non 
overlapping subgraphs -- the ``sectors'' -- such that all the external points of 
$X_i$ belong to the same sector, called an external sector.
In general, there exist sectors not containing external points, called internal 
sectors. Internal points may belong to external as well as to internal sectors, 
depending on the partition considered.

$\bullet$  {\em{Sector numbers}}: To any sector $S$, a number $s(S)$ is assigned 
according to the following rules.

(i) For the sector containing the external points $X_i$:  $s = i$.

(ii) For an internal sector $S$, $s(S)$ is a non-integer number between the maximum 
and the minimum sector numbers relative to the neighbouring sectors 
({\em i.e.} the sectors connected to $S$ by at least by one line of the graph).

(iii) If $\sigma_i \ne \sigma_{i+1}$ there is no internal sector with $i<s(S)<i+1$.

$\bullet$ {\em{Equivalence}}: If two partitions only differ in the numbering of the 
sectors -- not in their topology -- they are inequivalent if for at least one pair 
of neighbouring sectors $S^\prime,\,S^{\prime\prime}$ one has 
$s(S^\prime)>s(S^{\prime\prime})$ in the first
partition, $s(S^\prime)<s(S^{\prime\prime})$ in the second.

$\bullet$ {\em{Type}}: The sectors are either $T^+$ or $T^-$ sectors in the following 
way:
the external sector with number $i$ is a $T^{\sigma_i}$ sector; the internal 
sector with $i<s(S)<i+1$ and $\sigma_i = \sigma_{i+1}$ is a  $T^{\sigma_i}$ 
sector as well.

$\bullet$ {\em{Diagrammatic rules}}: Any partition is converted into an analytical 
expression according to the following.

(i) Inside a  $T^+$ sector ordinary Feynman rules for propagator and vertices 
apply.

(ii) Inside a $T^-$ sector the complex conjugate of Feynman rules hold.

(iii) Any internal sector contributes a $(-1)$ factor.

(iv) Finally, a line connecting two different sectors $S^\prime$ and $S^{\prime\prime}$  
corresponds, in momentum space, to a factor 
\begin{eqnarray}
\delta_{ij}\,(\bslp + \mu) \, 2\pi\,\theta(\pm p_0) \,
\delta(p^2-\mu^2) \qquad {\rm quarks} \quad ,  \label{cutf}
\end{eqnarray}
\begin{eqnarray}
\delta_{ab}\,\Bigl( -g_{\mu\nu} \, 2\pi\,\theta(\pm k_0)\,
\delta(k^2) + k_\mu k_\nu \cdots \Bigr )  \quad {\rm gluons}    \quad ,
\label{cutb}
\end{eqnarray}
where the dots in the second stand for gauge terms that decouple in all the 
$W$ functions we will calculate and the $\pm$ applies according to whether the 
number sectors satisfy $s(S^\prime)\lg s(S^{\prime\prime})$.

$\bullet$ {\em{Sum}}: The contribution of order $g^N$ to $W$ is obtained by summing 
the contribution of all inequivalent partitions so obtained and multiplying 
the result for the appropriate combinatorial factor. 

The inclusion of the case $T^0=1$ of no ordering  is taken into 
account by the  following observation.
Single fields $\Phi(x)$ are included in the above scheme by allowing external 
sectors with only one field as argument:  $\Phi(x)=T^\pm\bigl (\Phi(x)\bigr )$. In this way the single partitions of a graph do
depend on the choice of the sign, but the sum, expectedly, does not. This, 
in particular, provides the algorithm for computing Wightman functions in PT.

Some comments about the above Steinmann rules are in order. The iterative derivation of the above rules is based on the following inputs: 

(i) the equations of motion of the model; 

(ii) Wightman axioms
for the Wightman functions (including locality, but excluding positivity); 

(iii) on-shell normalization
conditions.

Within these assumptions the solution provided by the above rules is shown to 
be unique.

Concerning the last point we emphasize that, whenever needed, an IR regulator 
must be at work (which one is suitable for the models considered here will be 
discussed later). 

Moreover Steinmann himself emphasizes that no use of the asymptotic condition 
is ever made. This is quite welcome for us for, in the contrary case, this would imply some 
assumption on the IR asymptotic dynamics: this is exactly what we 
do not want to do.

The above rules provide the tool necessary for computing in PT, at least in 
principle, all the correlation functions  of the gauge invariant charged 
fields, as the ``quark'' $(\ref{PsiPT})$: this algorithm provides us immediately 
with  the ``quantum part'' of the calculation, {\em i.e.} that part that only  
involves the quantum fields in the r.h.s of $(\ref{PsiPT})$. 
About this part one should also observe that all the degrees of freedom, 
physical as well as unphysical, are associated to local fields that propagate 
in causal way.

However, there remains the ``classical part'' of the calculation, consisting in 
checking whether the $d^4k_j$ integrations involving both the VEV's and
the $\hat f$'s we have chosen 
(that should provide the decoupling of the unphysical degrees of freedom) 
are well defined.

We face this problem in the next section where we only consider two-point 
functions, because the rules we have reported above are somewhat unusual and 
more complicated than the Feynman rules everybody is used to: we better start 
learning the new game in the simplest case.


\section{Two-point Functions}    \label{sec4}

Our aim is to see how the algorithm given in the preceding section works in 
the case of the two-point function 
\widetext
\begin{eqnarray}
                &&  \int d^4 x \,e^{i\,p\cdot x}~ 
                    \int d^4 y \,e^{i\,q\cdot y}~
                    \langle \Psi(x;m)\overline \Psi(y;n)\rangle 
                  = (2\pi)^4 \delta_4(p+q)~W(p,m;q,n) =                                        \nonumber \\              
               && = \sum_{M,N=0}^\infty \quad \sum_{\sigma,\tau=\pm,0} (-ig)^M(ig)^N\,
                    \prod_{i=1}^M \int {d^4 k_i   \over (2\pi)^4}~ 
                    \prod_{j=1}^N \int {d^4 \ell_j\over (2\pi)^4}   \quad \times               \label{Wpmqn} \\    
               && \times ~~{\zeta_\sigma^{1/2}} \, \langle V^\dagger_\sigma\rangle^{-1} \,
                                \chi_\sigma(m)\,\hat f_{\sigma\,M}^{\mu_1\cdots k_M}(k_1\cdots,k_M;m) ~~~~
                           {\zeta_\tau^{1/2}}    \,\langle V_\tau \rangle^{-1} \,
                                \chi_\tau(n)\, \hat f_{\tau\,N}^{\nu_1\cdots \nu_N}(\ell_1\cdots,\ell_N;n)~\times      \nonumber \\
              && \times~~ \bigl \langle T^\sigma  
                                                \bigl [ \hat {\bf A}_{\mu_M}(k_M) \cdots 
                                                        \hat {\bf A}_{\mu_1}(k_1)\,\hat \psi \bigl (p-\sum k_i\bigr) \bigr]\,
                                        T^\tau  \bigl [ \hat {\overline \psi}(q-\sum \ell_j)\, 
                                                        \hat {\bf A}_{\nu_1}(\ell_1)\cdots \hat {\bf A}_{\nu_N}(\ell_N) \bigr ]\,
                          \bigr \rangle \quad .   \nonumber 
\end{eqnarray}
\narrowtext
Due to the presence of the string vectors $m$ and $n$, this two-point function extends 
equation $(\ref{WQdef})$ that will be recovered in the end of this section. 
In analogy to $(\ref{wdef})$ we will denote the amputation of 
$(\ref{Wpmqn})$ by
\begin{eqnarray}
               (2\pi)^4\delta_4(p+q)\,w(p,m;q,n) =\gamma_\alpha\,t^a\,\bigl 
                                                  \langle Q^{a\alpha}(p,m)~ \overline Q^{\,b\beta}(q,n)\bigr \rangle\,\gamma_\beta \, t^b \quad ,
\end{eqnarray}
where the $Q$'s are the currents defined in $(\ref{mDirac})$. 

Consistently with Steinmann assumptions, we assume that the QCD 
Lagrangian $\cite{NO}$ has been IR regulated and
renormalized with on-shell normalization conditions. 

Up to order $g^2$ the calculation is essentially abelian: 
the colour matrices in the two vertices contract to $C_{_F}$($\to 1$ for QED) 
and there is no three-gluon vertex. 
To this order, therefore, one can think of regularizing IR divergences by 
giving a mass $\lambda$ to the photon/gluon and UV
divergences by dimensional regularization: 
$4 \to 4 - 2\,\varepsilon,\,\varepsilon>0$.

As a matter of fact, on going to order $g^4$ it will be seen in $\cite{dEMi}$ that 
the mass regularization is not adequate and we shall use dimensional 
regularization $4 \to 4 + 2\,\epsilon,\,\epsilon>0$ for the IR $\cite{GM,MS}$
(this IR $\epsilon$ should not be 
confused with the UV $\varepsilon$, anyway they will never be 
simultaneously used)  and non-lagrangian Pauli-Villars $\cite{BD}$ for UV. 
Details about the problems connected with the choice of the regularizations are given in 
Section \ref{sec6}.

It is convenient to group the graphs contributing to the VEV in 
$(\ref{Wpmqn})$ in the following way:

(1) Usual or local graphs: those with the $M=N=0$ in the above double series, {\em i.e.}
the graphs contributing to the Wightman function 
$\langle \psi(x)\overline\psi(y)\rangle$.

(2L) Left graphs: $M>0,\,N=0$.

(2R) Right graphs: $M=0,\,N>0$, specular to the left graphs.

(3) Left/Right graphs: both $M>0$ and $N>0$.

This is exemplified  by the four graphs in FIG. $\ref{fig: Fig1}$, that gives the graphs 
contributing to order $g^2$. 

The sector partitions of the above graphs depend on whether either 
$m$ or $n$  are chosen in ${\cal C}_\pm$ or in the complement ${\cal C}_0$ 
of the light cone.  
To cover all the nine possibilities, it would be sufficient to consider only 
five cases, thanks to the Dirac conjugation properties of the fermion field, equation
$(\ref{DiracConjx})$. However, we will furtherly  restrict ourselves 
only to the three cases that are more interesting for our purposes:

(A) $m\in{\cal C}_+$, $n\in{\cal C}_-$;

(B) $m\in{\cal C}_+$, $n\in {\cal C}_0$;

(C) $m,\,n\in {\cal C}_0$.

The discussion of the remaining cases is, after these, a simple exercise.

In any event, the lowest order graph, common to all cases, contributes the 
free two-point Wightman function of the spinor field, equation $(\ref{Wfree})$. 

\subsection{$m\in{\cal C}_+$, $n\in{\cal C}_-$}    \label{sec4a}

Only the term of $(\ref{Wpmqn})$ with $\sigma = +,\,\tau =-$ contributes,
there are only two external sectors, sector 1 in the right that is a $T^-$ 
sector, and sector 2 in the left, that is a $T^+$ sector.
Since the two sectors are of different type, there can be no internal 
sectors.
The partitions of the above six graphs are thus obtained by drawing a 
cutting vertical line in all possible positions. 
In the cut lines we convene that momentum always flows from right to left, 
{\em i.e.} from sector 1 to sector 2 so that the replacement rules 
$(\ref{cutf})$ and $(\ref{cutb})$ are always taken with the plus sign.

All this resembles, and is nothing else but, the familiar Cutkosky-Veltman 
cutting rules. It should be noted that this regards only the VEV in the last line of 
$(\ref{Wpmqn})$. The $\hat f$-vertices contributed by the string operators, not even drawn in 
FIG. $\ref{fig: Fig1}$, are not touched upon by Steinmann rules: their denominators are 
instead prescribed by our definition $(\ref{PsiPT})$.

In addition, this identification of Steinmann rules with Cutkosky-Veltman 
rules happens only thanks to the choice made for $m$ and $n$. 
Different choices, as well as  VEV's with more
than two external sectors, are covered only by Steinmann rules.

The partitions drawn in FIG. $\ref{fig: Fig2}$-$\ref{fig: Fig4}$ refer to $(\ref{Wpmqn})$,
{\em i.e.} to the whole $\langle \Psi \overline \Psi \rangle$, 
not only to the VEV in $(\ref{PsiPT})$: the vertical lines represent the  
string denominators of the $\hat f$'s, whereas each vertex on a vertical 
line -- an empty circle -- contributes a factor proportional to  either 
$ g\, m_\mu $ or $ g\, n_\nu $; 
there also is a 4-dimensional integration for the loop.

Concerning graph A, its partitions are given in FIG. $\ref{fig: Fig2}$. 
Only the one marked {\tt (a)} is non-zero, the other
two vanish thanks to both mass and wave function
on-shell normalization conditions.

Graph BL has the two partitions given in FIG. $\ref{fig: Fig3}$ and named {\tt (bL)} and {\tt (}$\zeta${\tt L)}.
There are the two specular and
complex conjugate partitions {\tt (bR)} and  {\tt(}$\zeta${\tt R)} from BR.

Graph C has the only partition {\tt (c)} given in FIG. $\ref{fig: Fig4}$.

Graph TL too has only the partition {\tt (tL)} given in FIG. $\ref{fig: Fig5}$. 
There also is the partition {\tt (tR)} complex conjugate of the above.

We start with discussing the last graph.
It is ill-defined because its contribution to $W_1(p,m;q,n)$ is proportional 
to the integral
\begin{eqnarray}
  \int d_4 k~{{ g~m^\mu } \over{m \cdot k - i \,0}}~
       {{ g~m^\nu } \over{m \cdot (k - k) - i \,0}}~
       {{- i\, g _{\mu\nu} + \cdots} \over{k^2 - \lambda^2 + i\, 0}} 
\label{4.3}
\end{eqnarray}
that is not defined. Even in QED, where, due to absence of colour matrices, 
one could take for $\hat{f}_2$ the symmetrized form
\begin{eqnarray}
                \hat{f}_{+2}^{\mu \nu}(k_1,k_2;m) =  
{1 \over 2!}~{i\,m_\mu \over {m\cdot k_1 - i \,0}}~
{i\,m_\nu \over{m \cdot k_2 - i \,0}}~, \nonumber
\end{eqnarray}
the momentum conservation $k_1 = - k_2$ from the photon propagator would 
yield the integral 
\begin{eqnarray}
                \int d_4 k~{1 \over {k^2 - \lambda^2 + i \,0}}~
{1 \over {m \cdot k - i \,0}}~{1 \over {m\cdot k + i \,0}} 
\nonumber
\end{eqnarray}
plagued with a pinch singularity. So one has to get rid of it. 
This is exactly the task of the
factors $\langle V^\dagger_\sigma \rangle^{-1}$ in $(\ref{Wpmqn})$, as we 
now explain. 

The initial observation is that, thanks to translation invariance, the VEV 
of $V(x;m)$ cannot depend on $x$. So it may only
be a (ill-defined) constant times the identity matrix in colour space.
Imagine now that the theory has been
provisionally regularized by defining it on a space-time of finite volume 
$\Omega$: translation invariance is temporarily broken and momentum conservation 
does not hold, so that $(\ref{4.3})$ is now well defined: all the graphs depend
on $\Omega$ and tend to the expression that the above rules provide for them in the limit
$\Omega \to \infty$. However, before the limit is taken and up to order 
$g^2$, the factor  $\langle V^\dagger_+ \rangle^{-1}$ times $W_0(p)$ 
provides exactly the partition {\tt (tL)}, but with opposite sign.

Independently of any heuristic explanation, the factors 
$\langle V^\dagger_\sigma \rangle^{-1}$ are the instruction for the neglect 
of all the graphs including self interaction of the strings, as that
given in FIG. $\ref{fig: Fig8}$,
{\em i.e.} of the graphs that can be disconnected with one only cut in the 
string associated with the field $\Psi(x;m)$ -- one could call them One String Reducible graphs.

Likewise, $\langle V_\sigma \rangle^{-1}$ operates on the One String 
Reducible graphs associated with the string of the field 
$\overline{\Psi}(y;n)$.

We thus arrive to the conclusion that to order $g^2$ only the six partitions   
{\tt (a)}, {\tt (bL)}, {\tt (bR)}, {\tt (c)} and {\tt(}$\zeta${\tt L)}, {\tt(}$\zeta${\tt R)} survive, as well
as the counterterms coming from the expansion of 
$\zeta_+ = \zeta_- \simeq 1 + {\alpha/\pi}\, \zeta_1$.
For example the partition {\tt(}$\zeta${\tt L)} can be parameterized in the form:
\begin{eqnarray}
      {\tt (}\zeta {\tt L)} = {\alpha\over  \pi}\, C_{_F}\, 
      \left [ \gamma(\beta(m,p);{\rm UV},{\rm IR}) + \delta (\beta(m,p))\,
      {\bslm\,\mu \over m \cdot p} \right ]\,W_0(p) \label{zetal}
\end {eqnarray}
where, in terms of the ultraviolet and infra-red cutoffs 
\begin{eqnarray}
       && {\rm UV} = {1\over \varepsilon} - \gamma_{_{E}} + \ln{ 4\pi \kappa^2 \over \mu^2} 
                     \hspace{3.0 cm} {\rm{\small{dimensional~regularization}}}~,   \label{UV}  \\
       && {\rm IR} = \ln {\lambda \over \mu} 
                      \hspace{5.6 cm} {\rm{\small{mass~term~for~the~vector~meson}}}~, \label{IR}
\end{eqnarray}
and of the functions
\begin{eqnarray}
       &&  \beta(m,p) = \sqrt{ 1-{ {m^2\,p^2} \over (m\cdot p)^2} }~,       \quad
                        m,\,p\in {\cal C}_\pm  \Rightarrow  0<\beta<1 ~,    \quad
                        m \to p  \Rightarrow  \beta \to 0  ~,                 \label{beta} \\
       &&  B(\beta) = {1 \over 2\, \beta}\, 
                     \ln \,\left \vert {1+\beta \over 1-\beta}\right \vert ~, \label{B} \\
       &&  \Xi(\beta) = 
                       \left [{1\over \beta} \,{\rm Li}_2(\beta) 
                       +{1\over 2\beta}\, {\rm Li}_2\left (-{1+\beta\over 1-\beta}\right ) \right ] +
                        [\beta \to - \beta] \quad ,  \label{Xi}
\end{eqnarray}
the calculation of the  invariant functions $\gamma$ and $\delta$ gives the result:
\begin{eqnarray}
              && \gamma(\beta;{\rm UV}, {\rm IR}) = {1 \over 2}\,
                                                  \Biggl \{ {1\over 2}\,{\rm UV} + 1 + B(\beta)\,
                                                               \left ( 2 \,{\rm IR} + \ln{1-\beta^2\over 4} + 1 
                                                               \right )-~\Xi(\beta) + \label{gamma} \\
              &&  \hspace{3.2 cm}  +\left ( {1\over \xi}-1\right ) 
                                                           \left ({1\over 4}\, {\rm UV} + {\rm IR} + {7\over 4} - 
                                                                 {1\over 2}\,{\ln \xi\over 1-\xi} \right)
                                                  \Biggr \}\,W_0(p) ~ , \nonumber      \\         
             && \delta (\beta)  = - {1\over 2}\,B(\beta) \quad . \label{delta}
\end {eqnarray}

In  $(\ref{gamma})$ the contribution 
of the $g_{\mu\nu}$ and of the longitudinal terms of the vector meson 
propagator are the first and the second line, respectively.

Likewise, the calculation of {\tt(}$\zeta${\tt R)} is obtained by $(\ref{zetal})$ with the 
replacement:
\begin{eqnarray}
                 {\tt(}\zeta {\tt R)} = {\alpha \over \pi}\, C_{_F}\,W_0(p) \,\Bigl [ m \to n \Bigr ] \quad .    \label{zetar}
\end{eqnarray}

Obviously, the choice of $\zeta_1$ can only 
modify the invariant function $\gamma$. 

The first thing to note about the above graphs is that the coefficient of 
UV in $\gamma$ does not depend either on $p$ or on $m,\,n$. Therefore, this 
dependence (as well as the dependence on $\xi$) can be renormalized away.

The second thing to note is that the coefficient of IR -- proportional to 
$B(\beta(m,p))+B(\beta(n,p))$ -- does depend on $p$: 
the infra-red divergence cannot be eliminated by renormalization.

We choose
\begin{eqnarray}
                 {\alpha\over \pi}\,\zeta_1 = {\alpha\over \pi}\,C_{_F}\,
                 \left \{-{1\over 2}\,{\rm UV} - 2 \, {\rm IR} + 1 
                 + \left ( {1\over \xi}-1\right )\left ({1\over 4}\, {\rm UV} + {\rm IR} + 
                 {7\over 4} - {1\over 2}\,{\ln \xi\over 1-\xi}\right)\right \}~,
\end{eqnarray}
in which the finite part of $\zeta_1$  has been chosen in such a way that when both 
the limits \mbox{$m\to p,\,n\to -p$} are taken in $(\ref{zetal})$ and 
$(\ref{zetar})$ respectively, one obtains
\begin{eqnarray}
    {\tt(}\zeta  {\tt L)} + {\tt(}\zeta {\tt R)} + 
    {\alpha\over \pi}\,\zeta_1\, W_0(p) ~ 
    \stackrel {m,-n\,\to\,p}{\longrightarrow} 0~.
\label{lrz}
\end{eqnarray}

We have now to discuss the sector partitions {\tt (a)}, {\tt (bL)}, {\tt (bR)}, {\tt (c)}. 
They have in common the 2-body phase-space
\begin{eqnarray}
             & \Gamma_2(p)  &= \int d\Gamma_2 = \int \! d^4 k\,\theta(k_0)\,\delta(k^2)\,\theta(p_0-k_0) \, \delta((p-k)^2-\mu^2) =  \nonumber \\
             & &              = {\pi\over 2}\,\theta(p_0)\,\theta(p^2-\mu^2)\,\left (1-{\mu^2\over p^2}\right )         \label{2bps}
\end{eqnarray}
and their contribution to the amputated two-point function $(\ref{wdef})$ is
\begin{eqnarray}
                && {w}_1(p,m;-p,n) = {g^2 \over (2\pi)^2}\,C_{_F}\, \int d \Gamma_2~N(p;m,n)       \label{dgammaN}
\end{eqnarray}
where from
\begin{eqnarray}
       N(p;m,n) = 
       \left[ \gamma^\mu - (\bslp - \mu) {m^\mu \over {m \cdot k}} \right]
       (\bslp - \bslk + \mu)            
       \left[ \gamma^\nu - (\bslp - \mu) {n^\nu \over {n \cdot k}} \right]
       (- g_{\mu \nu} )            \label{N}
\end{eqnarray}
the contribution of each sector partition is clearly identifiable.

The following comments should help.

(a) The factors $(\bslp - \mu)$ in the square brackets of $(\ref{N})$ 
are due to the amputation.

(b)  The contribution of the spurious degrees of freedom in the gluon 
propagator is obtained by the replacements 
\[   \delta(k^2)\to \delta(k^2-\lambda^2) - \delta(k^2-\lambda^2/\xi)      \]
in the two-body phase-space $(\ref{2bps})$ and  
\[   - g_{\mu \nu} \to k_\mu k_\nu/\lambda^2                               \]
in $(\ref{N})$. The latter converts each of the square brackets into  
$(\bslp \,- \bslk - \mu)$ that in turn, on multiplying the factor 
$(\bslp \,- \bslk + \mu)$, gives
zero -- thanks to the delta function in the fermion phase-space.

(c) The prescriptions $\pm i\,0$ in the string vertex denominators have been
omitted inasmuch  as irrelevant to $(\ref{N})$: indeed, $m$, $-n$ and also 
$k$, thanks to  $(\ref{2bps})$, belong to ${\cal C}_+$, 
so that both $m \cdot k$ and $-n \cdot k$ are strictly positive on the 
two-body phase-space.

(d) For the same reason there is no need of IR regularization in 
$(\ref{dgammaN})$.

Use of covariance shows that $w_1(p,m;-p,n)$ can be expressed in terms of 
three integrals: one is $\Gamma_2(p)$, equation $(\ref{2bps})$. As for the others, 
it is convenient for later use to define 
\begin{eqnarray}
                I(p;m) = \int d\Gamma_2\,{\Pi(m)\over(m\cdot k)} ~ ,  \label{Idef} 
\end{eqnarray}
where $\Pi(m)$ is a prescription:
\begin{eqnarray}
  \Pi(m) =  \left \{ \begin{array}{lll}
                                          \pm 1 & {\rm if} & m\in {\cal C}_{\pm} \\  
                                          {\rm PV} &  {\rm if} &  m\in {\cal C}_0
                 \end{array} \right. ~,   \label{Pi}
\end{eqnarray}
and likewise
\begin{eqnarray}
                J(p;m,n) = \int d\Gamma_2\,{\Pi(m)\over (m\cdot k)}\,
                           {\Pi(n) \over (n\cdot k)} ~ . \label{Jdef}
\end{eqnarray}
The results of the calculations, for any $m$ and $n$, are
\begin{eqnarray}      
             && I(p,m)=\Gamma_2(p)\, {2\over p^2-\mu^2}\,
                        {p^2\over \vert p\cdot m \vert }\,B( \beta(m,p) ) ~ ,          \label{Ires} \\
             && J(p;m,n)=\Gamma_2(p)\,\,{4\over (p^2-\mu^2)^2}\,
                          {p^2\over \vert m\cdot n \vert}\,B( \beta(m,n) )  ~ , \label{Jres}
\end{eqnarray}
with $B$ given by $(\ref{B})$.

Only one observation is relevant about the above integrals, namely that the
limits $m\to -n$, $m\to p$, $n\to-p$ exist, commute with one another and commute 
with the phase-space integration to give
\begin{eqnarray}
             &&  I(p;p) = \Gamma_2(p)\,{2\over p^2-\mu^2} ~ , \\
             &&  J(p;m,-m) = J(p;p,-p) = \Gamma_2(p)\,{4\over (p^2-\mu^2)^2} ~ .
\end{eqnarray}

As long as the contribution of sector 
partitions {\tt (a)}, {\tt (bL)}, {\tt (bR)} and {\tt (c)} is infra-red finite, 
the lesson to be learned from adding this to 
the contribution of partitions  {\tt(}$\zeta${\tt L)} and {\tt(}$\zeta${\tt R)} (given by $(\ref{zetal})$, 
$(\ref{zetar})$) is that the perturbative theory for the field $\Psi(x;m)$, 
$m\in {\cal C}_+$, and its Dirac conjugate is plagued with the same 
IR pathology as for the gauge dependent $\psi(x)$.

Should one stop here, nothing would have been gained.

The only way to get rid of the IR divergence given by sector partitions 
{\tt(}$\zeta${\tt L)} and {\tt(}$\zeta${\tt R)} is to take both the limits $m\to p$ {\em and} 
$n\to -p$. In this case, due to the last two formulae, the contribution of sector partitions {\tt (a)}, {\tt (bL)}, {\tt (bR)}, {\tt (c)}
simplifies to
\begin{eqnarray}
              w_1(p)
             =C_{_F}\,\theta(p_0)\,\theta(1 - \varrho) \,(1 - \varrho)
             \left ( \bslp \, {{1 + \varrho}\over 2} - \mu \right ) \label{w1}
\end{eqnarray}
where
\begin{eqnarray}
             \varrho = \mu^2/p^2  \label{rho} \quad ,
\end{eqnarray}
whence, on reinserting the external propagators omitted for the amputation, 
taking into account $(\ref{lrz})$ and setting $C_{_F}=1$, one obtains the 
$W_1(p)$ appearing in $(\ref{WEres})$ and given by
equations  ($\ref{W2par}$), ($\ref{W1}$).

This is the piece of evidence that we can give in this paper, working to 
order $g^2$, about the existence -- and, to some extent, the necessity -- of 
taking the limit $(\ref{ntop})$, discussed in the Section \ref{sec2}.

The extension of $(\ref{w1})$ to the region $0<p^2 < \mu^2$ is legitimate and  trivial.

Also the extension of $(\ref{w1})$ to the region $p^2 < 0$ is trivial -- 
it also gives zero. But in this case there is a problem  of consistency between 
this extension, on the one side, and the Steinmann rules and the 
limits  $m, \, -n\to p$ on the other side. In this region infact the taking 
of the limits requires that  $m$ and/or $n$ be space-like from the outset 
and this, in turn, changes the sector partitions contributing  to $w_1$. 
It is however plausible to expect that the naive extrapolation of $(\ref{w1})$ to $p^2<0$ 
is correct: indeed all the sector partitions, even when calculated with the Steinmann 
rules suited for $m$ and/or $n$ space-like, should display 
either a $\Gamma_2(p)$ or a $W_0(p)$ factor, as encountered in the present
section. If really so, setting $p^2<0$ gives zero, due to the support 
properties of these factors, and the limits $m,\,-n \to p$ are quite safe. 

We feel however that, in order to check the above mentioned consistency, 
the exposing of the results of the explicit  calculation is more convincing. 
Also because, should one be interested in the perturbative
theory of the fields with $m\ne p$, there arise some difficulties connected 
with renormalizability that are better explicitly inspected. 

This is dealt with in the next subsections.

\subsection{$m\in{\cal C}_0$, $n\in{\cal C}_-$}    \label{sec4b}

\begin{figure}
\begin{center}
              \includegraphics[ scale=.5 , angle = 270]{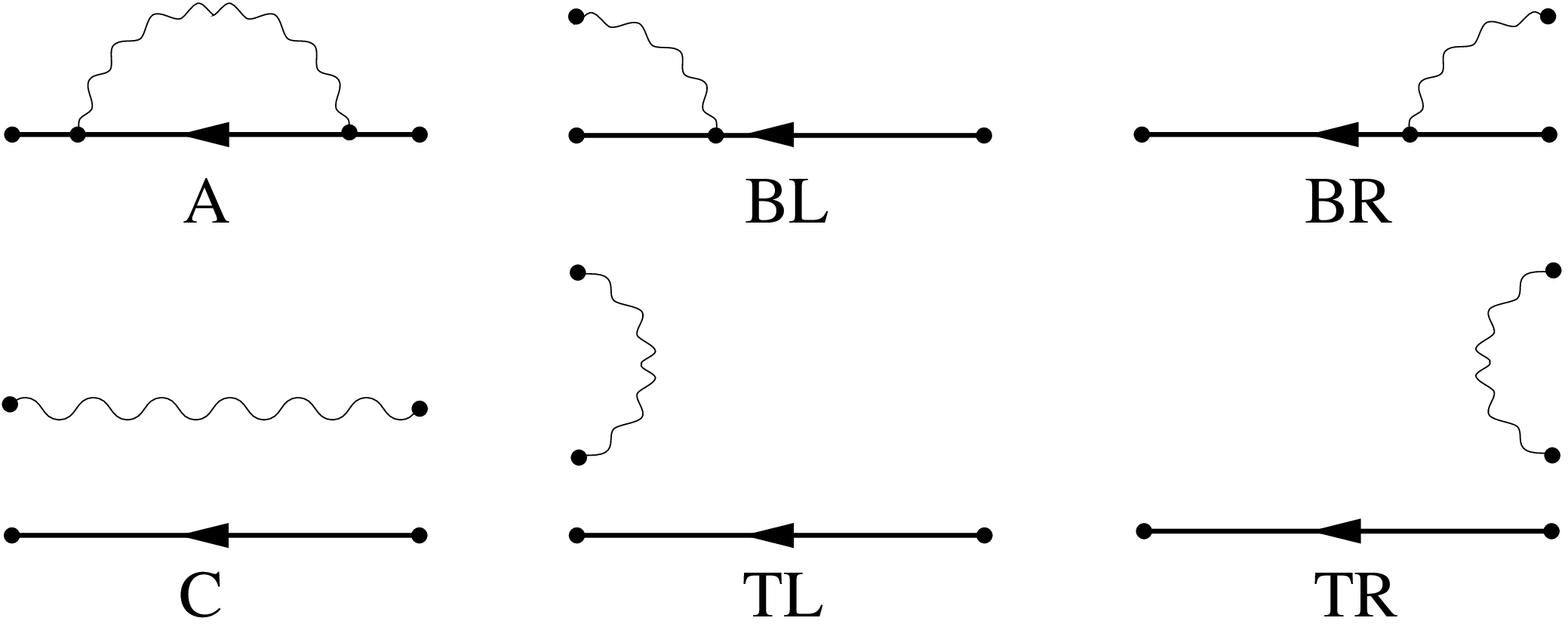}
              \vspace{.1 in}
\caption{~} \label{fig: Fig1}
\end{center}
\end{figure}

\begin{figure}
\begin{center}
              \includegraphics[ scale=.7 , angle = 270]{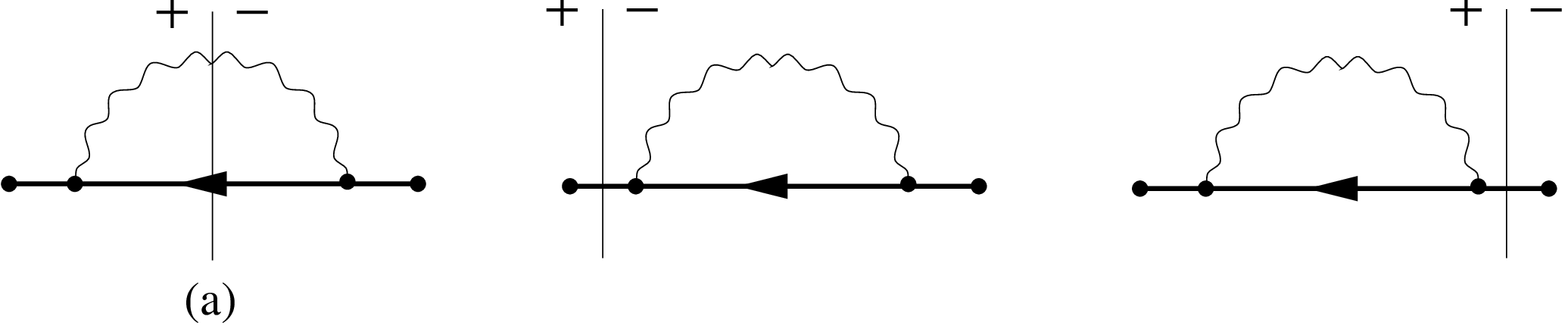}
\caption{~} \label{fig: Fig2}
\end{center}
\end{figure}

\begin{figure}
\begin{center}
              \includegraphics[ scale=.7 , angle = 270]{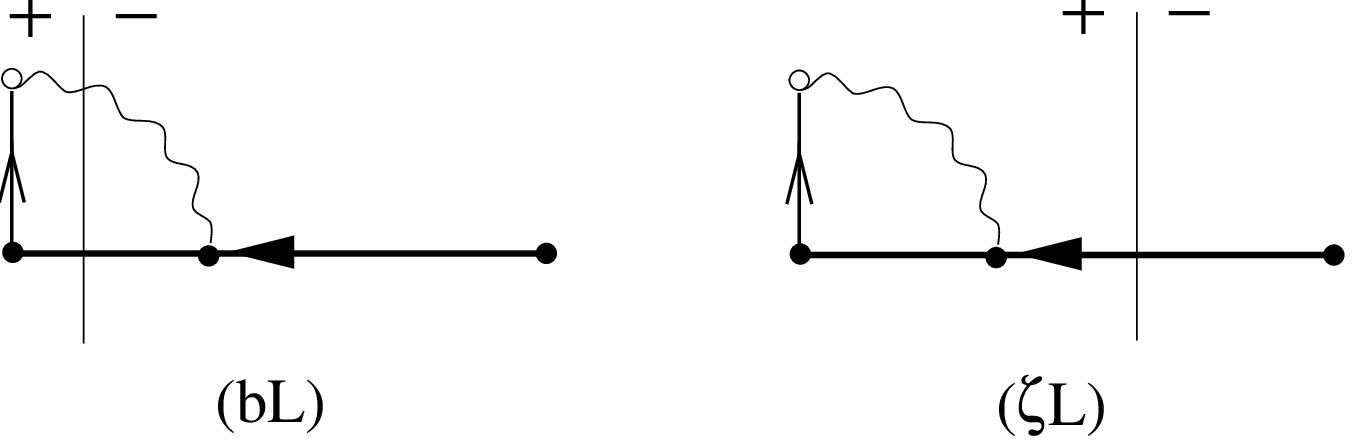}
              \vspace{.05 in}
\caption{~} \label{fig: Fig3}
\end{center}
\end{figure}

\begin{figure}
\begin{center}
              \includegraphics[ scale=.7 , angle = 270]{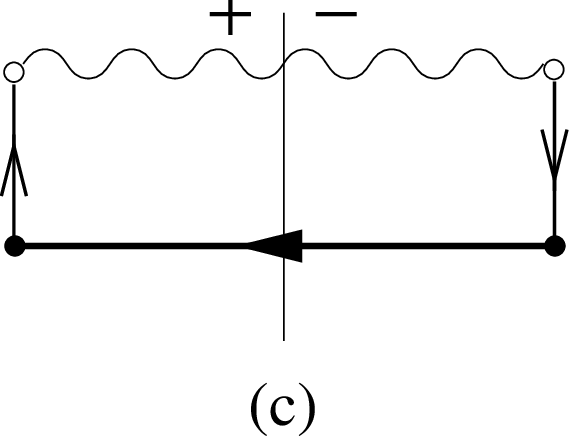}
              \vspace{.1 in}
\caption{~} \label{fig: Fig4}
\end{center}
\end{figure}

\begin{figure}
\begin{center}
              \includegraphics[ scale=.7 , angle = 270]{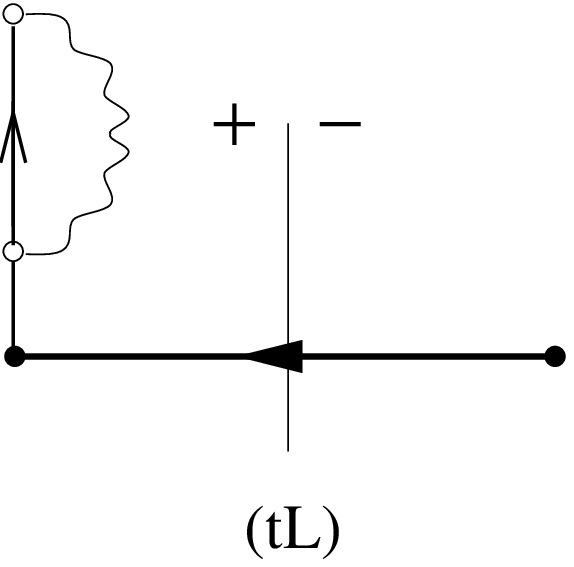}
              \vspace{.1 in}
\caption{~} \label{fig: Fig5}
\end{center}
\end{figure}

\begin{figure}
\begin{center}
              \includegraphics[ scale=.7 , angle = 270]{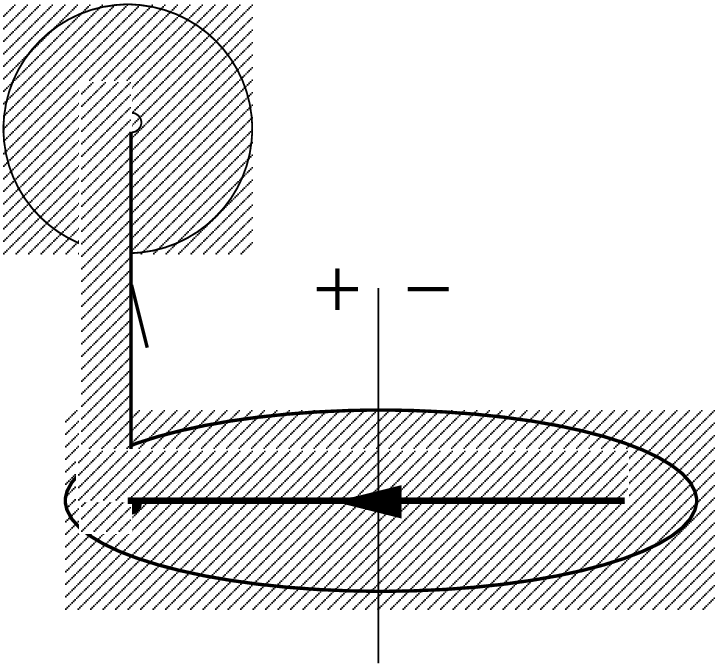}
               \vspace{.1 in}
\caption{~} \label{fig: Fig8}
\end{center}
\end{figure}

There is again the contribution of local graphs, namely those contributing to the 
ordinary Wightman function $\langle \psi(x)\overline \psi(y)\rangle$, {\em i.e.} graph 
A of FIG. $\ref{fig: Fig1}$. This is expected to be the same as in the previous 
section, as independent of the string vectors $m$ and $n$.
Indeed, as commented after the last Steinmann rule in Section \ref{sec3}, we have 
the freedom to assign a time-ordering label to each field, being sure that 
the final result does not depend on the assignment. 
We choose to write  
$\langle \psi(x)\overline \psi(y)\rangle=\langle T^+[\psi(x)]\,
T^-[\overline \psi(y)]\rangle$, 
that takes us back to the case discussed in the previous section: only 
sector partition {\tt (a)} of FIG. $\ref{fig: Fig2}$ gives a non-vanishing contribution.

\begin{figure}
\begin{center}
              \includegraphics[ scale=.7 , angle = 270]{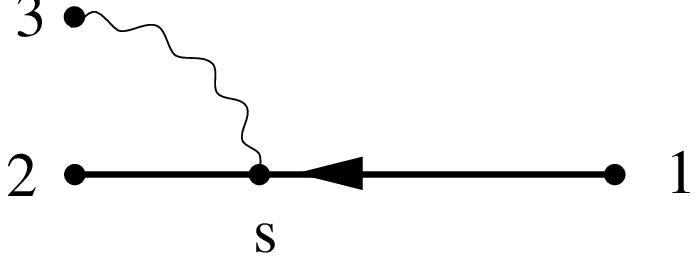}
\caption{~} \label{fig: Fig6}
\end{center}
\end{figure}

Let us now discuss the sector partitions of graph BL of FIG. $\ref{fig: Fig1}$. Now the three external vertices must be given sector numbers
as in FIG. $\ref{fig: Fig6}$
and the number $s$ can be given values $1 \le s \le 3$. 
So in principle there are five inequivalent partitions. 

The two partitions in which $s$ is non-integer have three on-shell lines 
joining in the same vertex, so their contribution is zero. 

There remain the three sector partition labelled by 
$s=1,\,2,\,3$.

The first -- $s=1$ -- is again {\tt (bL)} in FIG. $\ref{fig: Fig3}$, so its contribution is easily 
recovered from $(\ref{dgammaN})$ and $(\ref{N})$, provided the integral 
$(\ref{Idef})$ is taken, according to $(\ref{Wpmqn})$ 
and $(\ref{f10})$, with the ${\rm PV}$ prescription. 
In fact, in this case the denominator $m\cdot k$ is no longer positive on the
two-body phase-space. The result is still given by $(\ref{Ires})$.

\begin{figure}
\begin{center}
              \includegraphics[ scale=.7 , angle = 270]{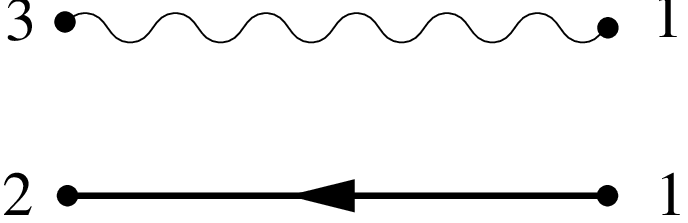}
\caption{~} \label{fig: Fig7}
\end{center}
\end{figure}

It is now convenient to consider the sector partitions of graph C 
in FIG. $\ref{fig: Fig1}$, postponing to later the
sector partitions of FIG. $\ref{fig: Fig6}$ labelled by $s$ = 2, 3. The sector numbers can only be assigned as in FIG. $\ref{fig: Fig7}$.
Therefore this is again partition {\tt (c)} of FIG. $\ref{fig: Fig4}$, easily recovered from
$(\ref{dgammaN})$ and $(\ref{N})$, provided that in the integral 
$(\ref{Jdef})$ the $m\cdot k$ denominator is PV prescribed.
Again the result is provided by  $(\ref{Jres})$.

Finally, the only  sector partition of the one string reducible graph 
TL in FIG. $\ref{fig: Fig1}$ (sector numbers 1 to 4 clockwise from right vertex) is again 
disposed of, thanks to $\langle V^\dagger_0 \rangle^{-1}$
instruction in $(\ref{Wpmqn})$. 

Going back to the other two sector partitions $s=2,\,3$ of FIG. $\ref{fig: Fig6}$,
they both have a $W_0(p)$ factor  (the fermion line in the right) and  so take 
the place of {\tt(}$\zeta${\tt L)} of the previous section. 

The contribution of these partitions is parameterized, in analogy with 
$(\ref{zetal})$, by 
\begin{eqnarray}
             {\tt{(zL)}} = {\alpha\over \pi}\, C_{_F}\,  \left [ c(\beta(m,p);{\rm UV},{\rm IR}) + d(\beta(m,p);{\rm UV})\,
                                                          {\bslm\,\mu\over m \cdot p} \right ]\,W_0(p)
\label{zl}
\end{eqnarray}
where, the invariant functions $c$ and $d$ depend on $m$ and $p$ through 
the variable $\beta(m,p)$ 
(defined in $(\ref{beta})$ and now, with $m \in {\cal C}_0$, satisfying $\beta > 1$) 
and on the cutoffs $(\ref{UV})$, $(\ref{IR})$: the result of the calculation gives
\begin{eqnarray}
             &c(\beta;{\rm UV},{\rm IR}) &= {\rm UV}\left ({1\over 2} +{3\over 8}\,\beta^{-2}+{1\over 8}(1-\beta^{-2})\,B(\beta) \right )
                                       +{\rm IR}\,B(\beta) + \nonumber \\
                             & &   +{5-\beta^{-2}\over 16}\,\Upsilon(\beta) 
                                       +{2\,\beta^{-2} -1\over 4}\,B(\beta) - {3\over 4}\,\beta^{-2} +\nonumber \\
                & & -{1\over 4}\left ({1\over \xi}-1\right) \left ( {\rm UV} - 2\,{\rm IR} +1\right ) + {1\over 4\, \xi}\,\ln \xi    \quad ,
\label{c} \\
            & d(\beta;{\rm UV}) &=  {\rm UV}\left (-{3\over 8}\,\beta^{-2} - {1\over 8}\,(1-\beta^{-2})\,B(\beta) \right ) + \nonumber\\
                                     & &+{1\over 4}\,\beta^{-2} - {1\over 4}\,B(\beta) -{1-\beta^{-2}\over 16}\,\Upsilon(\beta)  \quad ,
\label{d}
\end{eqnarray}
in which, we remind, $B(\beta)$ is given  by $(\ref{B})$ and 
\begin{eqnarray}
             \Upsilon(\beta) = \left [ {1\over 2\,\beta}\,\ln^2{1+\beta\over 2\, \beta} + {1\over \beta}\,
                                       {\rm Li}_2\left ({1+\beta\over 2\, \beta}\right ) \right ]
                                       + [\beta \to -\beta]     \quad .\label{Upsilon}
\end{eqnarray}

Comparison of the above formulae with the corresponding $(\ref{gamma})$, 
$(\ref{delta})$ shows  that -- contrary to $\delta$ -- the invariant function 
$d$ does depend on UV: this means that no choice of the renormalization 
constant $\zeta_0$ introduced in $(\ref{PsiPT})$ (and giving rise to 
counterterms proportional to $W_0(p)$, not to $\bslm  W_0(p)$) 
can cure the divergence. In addition there also are the coefficients of UV and IR in $c$ that
both depend on $p$ through $\beta$. As for the IR divergence associated
with a space-like string, the result is not new $\cite{S1}$.

There is a way out of this problem: choosing $m_0=0$ in the rest frame, in which only $p_0 = \mu\ne 0$. 
In this case in fact {\tt{(zL)}}$\to \alpha \, C_{_F}/\pi\, W_0(p)\cdot (UV/2\,+\,$last line of ($\ref{c}$)$)$, 
so that a suitable choice of $\zeta_0$ to order $g^2$ removes the divergence.

Unfortunately, the choice $m_0=0$ spoils Lorentz invariance and we will not stick 
to it.

To summarize: the perturbative theory involving a charged field, dressed 
with a string in space-like  direction, is non-renormalizable -- at least at 
finite orders -- due to the $m$-string. 
There survive, in addition, IR divergences carried by both the $m$- and the $n$-string.

\subsection{$m\in{\cal C}_0$, $n\in{\cal C}_0$}      \label{sec4c}

The discussion of local graphs as well as of the sector partitions with 
only one string vertex (FIG. $\ref{fig: Fig2}$; FIG. $\ref{fig: Fig6}$ and its analogue giving rise to
a partition {\tt (bR)} and to a contribution {\tt{(zR)}} obtained by ($\ref{zl}$), with a replacement analogue to ($\ref{zetar}$)) 
presents no novelty with respect to the preceding subsection.

The only novel feature is given by graph C of FIG. $\ref{fig: Fig1}$, where the only 
partition is given by assigning sector numbers from 1 to 4 with clockwise 
movement, starting from the top right vertex.

This is again recovered from $(\ref{dgammaN})$ and $(\ref{N})$, provided 
the integral $(\ref{Jdef})$ is now taken with both $m$ and $n$ space-like, 
{\em i.e} with both denominators prescribed by PV.
The result is once more $(\ref{Jres})$.

In this case also, thanks to the {\tt{(zL)}} and {\tt{(zR)}} contributions,
there are UV as well IR divergences due to each string. 

Choosing both $m_0=n_0=0$ in the rest frame would eliminate the two problems.

However, once more we refrain from breaking Lorentz symmetry, 
also because there is another way out of this empasse.
This is provided exactly by the double limit $m=-n \to p$. The latter has 
to be effected first by dragging $p$
in ${\cal C}_0$: this makes the whole two-point function vanish, 
due to the support of the $\delta(p^2-\mu^2)$ in
the {\tt{(zL)}} and {\tt{(zR)}} contributions, and to the support of the $\Gamma_2(p)$ 
two-body phase-space in all the other ones.
At this point taking the limit is safe and gives zero, in agreement with 
the naive extrapolation of $(\ref{w1})$ discussed in the end of 
subsection \ref{sec4a}.

\section{outlook of fourth order calculations}    \label{sec6}

The calculations of Section \ref{sec4} should make it evident that the
only two-point function free from both UV and IR problems is that
relative to the field ($\ref{defq}$): they provide evidence for the necessity,
rather than the possibility, of taking the limit $n\to p$, whose
meaning and implications we have discussed in the final part of Section \ref{sec2}.

It is also clear that, in order to obtain the result ($\ref{w1}$), 
commuting the limit $n\to p$ with the loop integration makes the
calculation by far simpler and that, consistently, only the 
diagrammatic rules of subsection \ref{sec4a} have to be used in order to
get a non-vanishing two-point function.

Exactly in this way, we have performed the two-loop calculation of
$W_2$, equation ($\ref{WEres}$), in QED. This receives contribution from 12
graphs for a total of 19 non-vanishing partitions, 10 of which
involve two-body phase-space, the other 9 involve three-body phase-space.
The graphs with only one external fermion line on shell -- as  the partitions
{\tt(}$\zeta${\tt L)} or {\tt(}$\zeta${\tt R)} of Section \ref{sec4a} -- have not been included in the counting,
because, much as in the case of ($\ref{lrz}$),
they can be renormalized away with a suitable choice of the
fourth order contribution $\zeta_2$ to the renormalization constant 
$\zeta_\pm$.

Several graphs exhibit an IR divergences proportional to $\ln \lambda$.
In this QED calculation, the photon mass regularization has been adopted, for it is well known
not to interfere with either BRST symmetry or unitarity.
Indeed, we have a diagrammatic ({\em i.e.}
without the need of analytic calculations) proof of the decoupling of
unphysical degrees of freedom, in the form of gauge-fixing 
parameter independence:
\begin{eqnarray}
            \xi \,{\partial \over \partial \xi}\, W_2 = 0
\label{xi}
\end{eqnarray}

Furthermore, we have found that the  Grammer-Yennie 
$\cite{GY}$ method of control of IR divergences can be extended, in a 
straightforward way, to the graphs that include the eikonal string vertices. 
This results in a diagrammatic proof (analogue to that of $\cite{KAdR}$)
of a complete cancellation between the  IR divergences coming from
the three-body cut graphs and the two-body cut graphs.

What is left is the explicit result of the calculation that we 
report below to give concreteness to what we have said, 
although in its full form it is not illuminating.
For $p^2<9\,\mu^2$ ({\em i.e.} omitting graphs involving
closed fermion loops, that are irrelevant for the near-mass-shell
asymptotics) the two structure functions, defined by ($\ref{W2par}$) with $i=2$, are given by
\begin{eqnarray}
              && a_2={\frac {\varrho \,\left( -39 - 82\,\varrho  + 37\,{{\varrho }^2}\right) }
                            {16\,\left( 1 - \varrho  \right) } } +
                     {\frac {{{\varrho }^3}}
                            {2\,\left( 1 - \varrho  \right) } }\,\ln (1 - {\sqrt{1 - \varrho }})\,
                                                                 \ln (1 + {\sqrt{1 - \varrho }}) + \nonumber \\
  &&\hspace{0.6 in}+ {\frac {{{\varrho }^2}}
                            {2\,{\sqrt{1 - \varrho }}} }\,\ln \,{\frac{1 + {\sqrt{1 - \varrho }}}
                          {1 - {\sqrt{1 - \varrho }}}} +                 
                    {\frac {\varrho \,\left( 3 - 31\,\varrho  + 2\,{{\varrho }^2} - 2\,{{\varrho }^3} + 2\,{{\varrho }^4} \right) }
                            {8\,{{\left( 1 - \varrho  \right) }^2}}} \,\ln \,\varrho +  \nonumber \\
  && \hspace{0.6 in}+ {\frac { \varrho \,\left( -2 + 5\,\varrho  - 5\,{{\varrho }^2} + 3\,{{\varrho }^3} - {{~{\varrho }}^4}+
                              ( - 7 - 2\,\varrho  - {{~{\varrho }}^2} + 2\,{{\varrho }^3}) \,\ln \,\varrho\right)  }
                            {4\,{{\left( 1 - \varrho  \right) }^2}}}\,\ln (1 - \varrho ) +   \nonumber \\      
  && \hspace{0.6 in}+ {\frac {{{\varrho }^2}\,\left( -1 + 2\,\varrho  \right) }
                            {8\,\left( 1 - \varrho  \right) }}\,\ln^2 \,\varrho +        
                     {\frac {\varrho \,\left( -5 - 6\,\varrho  + 3\,{{\varrho }^2} \right)}
                            {4\,{{\left( 1 - \varrho  \right) }^2}}}\,\left( {\rm Li}_2(\varrho )- {\rm Li}_2(1)\right)   \\
           && \nonumber \\
           && b_2=   {\frac {\varrho \,\left( 5 + \varrho  \right) \,\left( -9 + 2\,\varrho  \right) }
                            {8\,\left( 1 - \varrho  \right) }} -
                     {\frac {{{\varrho }^2}}
                            {2\,\left( 1 - \varrho  \right) }} \ln (1 - {\sqrt{1 - \varrho }})\,
                                                               \ln (1 + {\sqrt{1 - \varrho }}) +  \nonumber \\
 && \hspace{0.6 in}- {\frac {\,\varrho }
                           {2\,\sqrt{1 - \varrho}   }}  \,\ln {\frac {1 + {\sqrt{1 - \varrho }}}
                                                                      {1 - {\sqrt{1 - \varrho }}}}\, +                        
                    {\frac {\varrho \,\left( -2 - 13\,\varrho  + 6\,{{\varrho }^2} - 5\,{{\varrho }^3} + {{\varrho }^4} \right)  }
                           {4\,{{\left( 1 - \varrho  \right) }^2}}}\,\ln \varrho\, + \nonumber \\
 && \hspace{0.6 in}- {\frac {{{\varrho }^2}}
                           {8\,\left( 1 - \varrho  \right) }}\,{{\ln^2 \varrho }} \,+   
                    {\frac {   \varrho \,\left( 3\,\varrho  - 7\,{{\varrho }^2} + 5\,{{\varrho }^3} - {{~{\varrho }}^4} +
                               ( - 13 + 4\,\varrho  + {{~{\varrho }}^2} )\,\ln \varrho \right) }
                          {4\,{{\left( 1 - \varrho  \right) }^2}}}\,\ln (1 - \varrho )\, + \nonumber \\
 && \hspace{0.6 in}+ {\frac {\varrho \,\left( -13 + 2\,\varrho  + 3\,{{\varrho }^2}
                            \right)  }{4\,
                         {{\left( 1 - \varrho  \right) }^2}}}\,\left( 
                          {\rm  Li}_2(\varrho )-{\rm  Li}_2(1) \right)
\end{eqnarray}
whose asymptotic form for $\varrho=  \mu^2/p^2\to 1$ is equation ($\ref{W2}$). It should be also noted
that in the ultraviolet regime $p^2\to +\infty$, {\em i.e.} for $\varrho \to 0$,  $a_2$ and $b_2$ vanish respectively
as $(\ln p^2)/p^2$ and $1/p^2$. So, when dispersed in $p^2$, they need no subtraction.

Concerning the QCD counterpart of the above calculation, stated
in equation ($\ref{ede}$) of Section \ref{intro}, apart from the contribution
of the non-planar QED graphs (those where the colour matrices occur
in the sequence $t^at^bt^at^b=C^2_{_F}- {1\over 2}\,C_{_A}C_{_F}$), there are 15 more graphs giving rise to 23
partitions, 8 of them  involving two-body phase-space, the other 15 the 
three-body phase-space.

A due remark concerns the IR regularization: 
giving a mass to the gluon, even 
according to $\cite{CF}$, preserves BRST symmetry, but only formally preserves unitarity
in the limit $\lambda \to 0$.  
As a matter of fact we have verified that, in this limit,  the r.h.s. of ($\ref{xi}$) 
does not vanish. 
We have therefore abandoned this
regularization adopting dimensional regularization for the IR
$\cite{GM,MS}$  and changed dimensional regularization for the UV with
non-lagrangian Pauli-Villars $\cite{BD}$. 
With these regularizations, equation ($\ref{xi}$) indeed holds also in the 
nonabelian case. 

In writing equation ($\ref{ede}$), we have recalculated the abelian part (proportional to 
$C^2_{_F}$) with the new IR and UV regularizations, with the expected
result that the cancellation of IR divergences
holds also in the new scheme. These details, however,  
will be part of a forthcoming paper giving the details of the above described two-loop
calculation $\cite{dEMi}$. 

Concerning instead the non-abelian part, we can say,
referring to the factor ${11\over 6}$ appearing in ($\ref{ede}$), that ${5\over 6}$ comes from the sum of all
the graphs that include gluon self-energy corrections; a further 1 comes from the
$C_{_A}C_{_F}$ part of the non-planar abelian graphs. For the other graphs (whose 
sector partitions are, in some cases, IR divergent even as $1/\epsilon^2$, not just as $1/\epsilon$)
there is a complete 
cancellation between the two-body cut and the three-body cut contributions to each of them.

\section{Conclusions}    \label{sec5}

We have shown how to construct BRST invariant
composite fermion fields that carry the global quantum numbers of the
electron and of the quark in QED and QCD.
The construction consists in
dressing the ordinary Dirac field with a rectilinear string whose
space-time direction is characterized by a 4-vector that, provisionally,
breaks the Lorentz covariance properties of the field.
In perturbation theory the string generates new graphs characterized by the 
occurrence of eikonal vertices. These new vertices
require prescriptions (either $\pm i\, 0$ or PV) whose choice is uniquely 
dictated  by the Dirac conjugation properties of the field. 
Furthermore, after going in momentum representation, the 4-vector 
characterizing the string must be chosen proportional to the 4-momentum of the 
field.
This choice: 
(i) restores Lorentz, 
(ii) averts some IR as well as some nonrenormalizable UV divergences.
The second point indicates that, as a matter of fact, there is little choice.

The whole construction survives the check of a fourth order calculation
of the two-point function in PT, performed both in QED and QCD. 

If these fields are to survive further and more stringent
verifications,  one can conclude that global charges associated to a Gauss law
imply, for the fields carrying such charges, delocalization properties considerably more involved than the single
1-dimensional string in 3-space, somewhat popular in the literature: since the string is rectilinear in 4-momentum space, in coordinate representation
the fields rather appear spread out all over Minkowski space, exhibiting a kind of
candy-sugar structure.

The construction gives -- as an extra bonus -- different
results for the IR asymptotic dynamics of QED and QCD respectively. In particular,
it hints at a mechanism of confinement according to which
the quark so constructed seems to behave as a free field at any momentum
scale.

The construction presented raises several problems; but 
the algorithm we have given in this paper also provides the possibility to
face them. Among others, a few of them still involve two-point 
functions:
\begin{itemize}
\item Extending to any order in PT the above results about:  
      (i) IR cancellation in QED and   
      (ii) IR non-cancellation and factorization in QCD.

\item The verification that any gauge invariant coloured field, first of all 
the gluon, has the same behaviour as the quark.
\end{itemize}

Other problems pertain instead a study of correlation functions with more than 
two points:

\begin{itemize}
\item The verification that such fields do indeed carry the
expected global charges, {\em e.g.} that they satisfy in PT at least the
weak commutation relations
\[    
      \langle \, e(x)\,[ Q\,, \overline e(y)]\, \rangle = 
      \langle \,  e(x)\, \overline e(y) \,\rangle ~,
\]
{\em etc.}, where $\displaystyle{Q=\int d^3 x :\!\overline \psi\, \gamma_0 \, \psi \!:\!(x)}$
 is the electric charge (and the analogue in QCD).

\item A study of the VEV of the algebra of Lorentz generators (the boosts
in particular), in order to represent, in  explicit way, the mechanism that
prevents a unitary implementation of Lorentz symmetry in the charged
superselection sector generated by $e(x)$ 
-- or, in alternative, to show how this mechanism is evaded by the fields we have proposed.

\item A comparison of the present approach with the well established results
of QED such as those about inclusive cross section or the electron $g-2$ and, in general, the
impact of this construction -- if any -- on the $S$ matrix.

\item For QCD, the proof of the scenario we have hinted at in Section \ref{intro}, namely
that amplitudes involving gauge invariant coloured fields either vanish
or disconnect into the product of free two-point functions relative to 
coloured fields times an amplitude that only involves colour singlet fields.

\end{itemize}

These problems are already under our investigation and we will report
about them in the near future.

\acknowledgements

The authors are greatly indebted to Dr. M. Mintchev, Dr. G. Morchio and Dr. D. McMullan for extensive
discussions on these topics. Dr. Mintchev is also acknowledged for having carefully and thoroughly read the manuscript. 
E.d'E. is grateful to Dr. B.R. Webber and Prof. J.B. Griffiths for the warm 
hospitality at the Cavendish Laboratory, University of Cambridge (GB), and at the Department 
of Mathematical Sciences, Loughborough University (GB) respectively, where great part of this 
work was done. S.M. also wishes to thank Prof. J.B. Griffiths for the encouragement and patient
support during the preparation of this work.



\begin{references}
\bibitem{BD} 
A book where radiative corrections are dealt with by this method is
D. Bjorken, S.D. Drell, {\sl Relativistic Quantum Fields}, (McGraw--Hill, New York, 1964)
\end{references}
\end{document}